\documentclass[preprint]{emulateapj} 

\usepackage{natbib}
\usepackage{hyperref}
\hypersetup{colorlinks,citecolor=Blue,linkcolor=Red,urlcolor=Blue}
\usepackage{amsmath}
\usepackage{empheq}
\usepackage[usenames,dvipsnames]{color}
\allowdisplaybreaks 

\newcommand{\feone}{{ f_{e,2}^{(1)} }} 
\newcommand{\fetwo}{{ f_{e,2}^{(2)} }} 
 
\newcommand{\fione}{{ f_{i,2}^{(1)} }} 
\newcommand{\fitwo}{{ f_{i,2}^{(2)} }}

\newcommand{\feonej}{{ f_{e,j}^{(1)} }} 
\newcommand{\fetwoj}{{ f_{e,j}^{(2)} }} 
\newcommand{\fethreej}{{ f_{e,j}^{(3)} }} 
\newcommand{\fionej}{{ f_{i,j}^{(1)} }} 
\newcommand{\fitwoj}{{ f_{i,j}^{(2)} }} 
\newcommand{\fithreej}{{ f_{i,j}^{(3)} }} 
\newcommand{\feij}{{ f_{ei,j}}} 

\newcommand{\Feone}{{ F_e^{(1)} }} 
\newcommand{\Fetwofive}{{ F_{e_5}^{(2)} }} 
\newcommand{\Fetwotwo}{{ F_{e_2}^{(2)} }} 
\newcommand{\Fethree}{{ F_e^{(3)} }} 
\newcommand{\Fione}{{ F_i^{(1)} }} 
\newcommand{\Fitwo}{{ F_{i_2}^{(2)} }} 
\newcommand{\Fithree}{{ F_i^{(3)} }} 
\newcommand{\Fei}{{ F_{ei}}} 

\newcommand{\icarus}{Icarus}
\newcommand{\Poincare}{{Poincar$\acute{\rm{e}}$}}
\newcommand{\appropto}{\mathrel{\vcenter{\offinterlineskip\halign{\hfil$##$\cr\propto\cr\noalign{\kern2pt}\sim\cr\noalign{\kern-2pt}}}}}

\begin{document}
 
\title{Chaotic Disintegration of the Inner Solar System}  

\author{Konstantin Batygin$^{1}$, Alessandro Morbidelli$^{2}$, Mathew J. Holman$^3$} 
\affil{$^1$Division of Geological and Planetary Sciences, California Institute of Technology, 1200 E. California Blvd., Pasadena, CA 91125}
\affil{$^2$Departement Lagrange, Observatoire de la C${\mathrm{\hat{o}}}$te d'Azur, 06304 Nice, France}
\affil{$^3$Harvard-Smithsonian Center for Astrophysics, 60 Garden St., Cambridge, MA 02138}

\begin{abstract}
On timescales that greatly exceed an orbital period, typical planetary orbits evolve in a stochastic yet stable fashion. On even longer timescales, however, planetary orbits can spontaneously transition from bounded to unbound chaotic states. Large-scale instabilities associated with such behavior appear to play a dominant role in shaping the architectures of planetary systems, including our own. Here we show how such transitions are possible, focusing on the specific case of the long-term evolution of Mercury. We develop a simple analytical model for Mercury's dynamics and elucidate the origins of its short term stochastic behavior as well as of its sudden progression to unbounded chaos. Our model allows us to estimate the timescale on which this transition is likely to be triggered, i.e. the dynamical lifetime of the Solar System as we know it. The formulated theory is consistent with the results of numerical simulations and is broadly applicable to extrasolar planetary systems dominated by secular interactions. These results constitute a significant advancement in our understanding of the processes responsible for sculpting of the dynamical structures of generic planetary systems.
\end{abstract}

\section{Introduction}
The question of whether or not the Solar System is destined to gravitationally unravel dates back to the very origins of physical science and Newton himself (see \citealt{LaskarPoincare}). However, the realization that planetary orbits in the Solar System exhibit chaotic motion and can become violently unstable is quite recent, especially when compared to the age of the problem itself \citep{Laskar89,Quinn1991,SussmanWisdom1988,SussmanWisdom1992}. Indeed, historically the vast majority of research dedicated to this issue had been aimed at obtaining a proof of the Solar System's indefinite permanence\footnote{Detailed accounts of the problem's history can be found in works of \citet{Laskar96} and \citet{LaskarPoincare}, to which we direct the curious reader.}.

The first compelling affirmation of the Solar System's stability stemmed from the works of \citet{Lagrange} and \citet{Laplace72,Laplace75}, while further extensions to the resulting secular theory were devised as a consequence of the calculations of \citet{Poisson,Gauss1809,Adams1846} and \citet{LeVerrier1855} among others. Nevertheless, the proof of the general non-integrability of the gravitational three-body problem brought forth by \citet{Poincare92} directly disputed previous claims of orbital regularity \citep{Laskar96}. 

%Despite the aforementioned demonstration of the fragility inherent to the method of successive approximations when applied to long-term orbital evolution, fueled by the advancements in Hamiltonian perturbation theory of the mid-20th century, \citet{Kolmogorov,Arnold1963} and \citet{Moser} rigorously showed the existence of quasi-periodic orbits in the planetary N-body problem. Although these results sparked considerable discussion with regards to their implications for the Solar System (see \citealt{Laskar96}), a meaningful application of the famed KAM theorem to the Solar System would require the planetary masses to be smaller by many tens of orders of magnitude \citep{Arnold1961}. 

Despite \Poincare's demonstration of the fragility inherent to the method of successive approximations, \citet{Arnold1963} rigorously showed the existence of quasi-periodic orbits in the planetary N-body problem. However, a direct application of KAM theory \citep{Kolmogorov,Arnold1961,Moser} to the three-body problem placed the estimate of Jupiter's threshold mass (below which stability is ensured) at a similar order of magnitude as the mass of a Hydrogen atom \citep{Henon1966}. As a result of subsequent efforts, this estimate was refined to a value much closer to Jupiter's actual mass \citep{1997CMaPh.186..413C,2000CeMDA..78...47L}, although the implications of this result for the Solar System as a whole remain limited due to the restricted scope of the calculation.

Substantial breakthroughs in the evaluation of the Solar System's long-term fate irrevocably came as advances in computer technology allowed extensive numerical calculations to illustrate the chaotic nature of the orbits \citep{1986AJ.....92..176A,1987A&A...181..182C,SussmanWisdom1988,SussmanWisdom1992,Laskar89,2002MNRAS.336..483I}. Accordingly, over the last two decades, the possibility of large-scale instability, fostered by Mercury's acquisition of a nearly radial orbit, has been demonstrated by a variety of dynamical models \citep{Laskar94,Laskar08,BatLaugh}. 

An examination of the numerical results revealed that Mercury's orbital excitation is facilitated by its entrance into the so-called $\nu_5$ secular resonance \citep{Laskar08,BatLaugh} and that general relativistic effects play a crucial role in diminishing the chances of such an event. The final nail into the coffin of the belief in the Solar System's enduring stability was delivered by the study of \citet{Laskar09}, who performed a series of high precision N-body simulations that appraised the chances of Mercury's ejection from the Solar System within the Sun's remaining main sequence lifetime at $\sim 1\%$, and confirmed the existence of collisional trajectories among the terrestrial planets. 

Following the numerical demonstrations of the possible onset of large-scale instability, a number of authors have began re-examining Mercury's dynamics from a perturbative point of view. To this end, \citet{LithWu11} applied the asteroidal secular model of \citet{Sid1990} to Mercury and semi-analytically demonstrated that in addition to the chaotic secular angle identified by \citet{Laskar89} (see also \citealt{SussmanWisdom1992}), a substantial number of higher-order secular angles stochastically switch between circulation and libration, hinting at the overlap of numerous high-order secular resonances (see \citealt{Chirikov1979}) as the source of Mercury's irregular motion. Building on the work of \citet{LithWu11}, \citet{Boue} showed that in the vicinity of the $\nu_5$ secular resonance (an orbital state that Mercury currently does not occupy but can evolve into), Mercury's acquisition of a highly eccentric orbit can be understood within the framework of a simple one degree of freedom Hamiltonian. 

Despite previous efforts, a crucial aspect of Mercury's chaotic evolution remains elusive. Specifically, the physical process underlying Mercury's abrupt transition from a chaotic, yet stable state to a violently unstable state, as well as the characteristic timescale for this transition remain poorly understood from a theoretical point of view.

Let us elaborate. Chaotic decoherence in the inner Solar System occurs on a timescale that is short compared with the Sun's age. Specifically, the characteristic Lyapunov time (a time required for initially nearby chaotic orbits to diverge by a factor of $e$) of the inner Solar System is of order a few Myr \citep{Laskar89,Laskar96,SussmanWisdom1992}. This means that since the acquisition of its current orbital state, Mercury has had the opportunity to lose correlation with its own initial condition approximately one thousand times. Thus, if the Solar System has existed in a state of vigorous, yet bounded chaos\footnote{Here, an analogy with weather on Earth begs to be made: while weather itself is unpredictable over timescales longer than a few days, typical temperature variations are scarcely expected to ever exceed tens of degrees.} for billions of years, what triggers the catastrophic ejection of Mercury, observed in the numerical simulations? Answering this question is the primary aim of this work.

The implications of qualitatively understanding the Solar System's long-term chaotic behavior range far beyond the special case of Mercury. In particular, the characterization of the orbital distribution of extrasolar planets (see e.g. \citealt{Cumming} and the references therein) as well as recent progress on quantifying the early evolution of our own Solar System \citep{2005Natur.435..459T, 2008Icar..196..258L, 2011AJ....142..152L, 2010ApJ...716.1323B, 2011ApJ...738...13B, 2012AJ....144..117N} has shown that dynamical instabilities\footnote{Numerical experiments (e.g. \citealt{Raymond09a, Raymond09b}) show that the onset of planet-planet scattering exhibits similar characteristics as the ejection of Mercury from the Solar System.} play a critical role is sculpting the orbital architectures of generic planetary systems \citep{RasioFord96, FordRasio08, Chaterjee08, JuricTremaine, Raymond09a}. Consequently, the formulation of a tangible theory for the onset of dynamical instabilities would without a doubt, significantly advance our overall understanding of post-nebular dynamical evolution.
 
In an effort to construct a simple model for the chaotic evolution of Mercury, we mirror the works of \citet{LithWu11,Boue} and opt for a perturbative treatment of the dynamics. Although not as precise as direct numerical investigation, this approach is qualitatively more fruitful. Indeed, perturbative analysis of the Asteroid belt's dynamical structure has been immensely useful in elucidating the origins of chaos and the characteristic properties of stochastic evolution (see e.g. \citealt{Wisdom80, Wisdom83,Morby90a,Morby90b,Morby91a,Morby91b, HolmanMurray1996, MurrayHolman97,DavidMorbya,DavidMorbyb} and the references therein). Moreover, a perturbative study of the outer Solar System's dynamical structure has yielded important insights into the expected dynamical lifetime of the giant planets \citep{MurrayHolman99}.

As a starting point, we shall adopt the well-known Lagrange-Laplace theory (see \citealt{MD99,Morbybook}) and sequentially enhance the complexity of our model until the desired behavior is adequately represented. Because the focus of the study lies in understanding the underlying physical processes, comprehensibility will be favored at the expense of quantitative precision. The formulated theory will allow for a purely analytical estimation of the Lyapunov time, chaotic diffusion rate, and the timescale for transition to global instability. Indeed, computational resources are in principle unnecessary (although extremely helpful) in completing the calculations presented here.

The paper is structured as follows. In section 2, we construct a secular model that captures the bounded stochastic behavior of Mercury. In section 3, we use this model to identify the primary secular resonances responsible for driving chaotic motion. Subsequently, we analytically deduce Mercury's Lyapunov time and chaotic action diffusion coefficient. In section 4, we extend the secular model to encapsulate the abrupt transition from bounded to unbounded chaos and analytically calculate the characteristic timescale for the onset of the instability. Additionally, we briefly discuss the sensitivity of the model to the general relativistic correction and its effects on the system's long-term stability. We conclude and discuss our results in section 7. 

\section{A Perturbative Model for Mercury's Chaotic Motion}

In this section, we present a series of models, based on classical series expansions of the Hamiltonian (see. e.g. \citealt{1995CeMDA..62..193L,2000Icar..147..129E}), with the aim of finding the simplest model that captures Mercury's stochastic secular eccentricity and inclination dynamics. At this stage, we need not concern ourselves with the onset of instability (which we show below requires an additional degree of complexity). Instead, we begin by elucidating the origin of ``short-term" (i.e. multi-Myr) chaotic behavior.

\subsection{A Linear Integrable Model}
As already mentioned above, the first complete description of the Solar System's secular dynamics arose from the works of Lagrange and Laplace in the late 1700's. The Lagrange-Laplace secular theory is strictly periodic and therefore cannot yield chaotic motion. Still, it comprises a useful starting point for the discussion that follows.

Within the framework of secular theory, Keplerian motion is averaged over, and the interactions among planetary orbits simplify to that of gravitationally coupled massive wires \citep{MD99,Morbybook}. To leading order in the planetary masses, eccentricities and inclinations, the orbits behave as a series of linked harmonic oscillators whose equations of motion constitute an eigenvalue problem.

\begin{figure*}[t]
\centering
\includegraphics[width=1\textwidth]{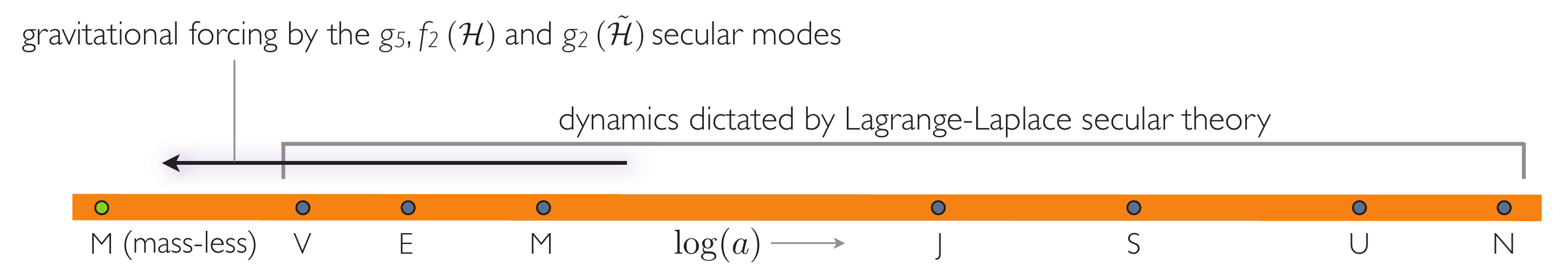}
\caption{Geometrical setup of the model system we address here. Mercury is treated as a test particle, and is gravitationally perturbed by the remaining planets. Keplerian motion is averaged over and the gravitational potential is expanded as a power series in the orbital eccentricities and inclinations. The strictly periodic Lagrange-Laplace secular solution is adopted as a description of the dynamical evolution of the perturbing planets. The $g_5$, $f_2$ and $g_2$ eigen-modes are exclusively retained in the disturbing function.}
\label{setup}
\end{figure*}

For clarity, let us begin with a setup relevant to the secular three body problem. Following, \citet{LithWu11} and \citet{Boue}, we first treat only perturbations due to Venus onto Mercury, assuming that Mercury provides no back-reaction onto Venus. In other words, we model the evolution of Mercury as that of a test-wire, subject to time-dependent external perturbations. Figure (\ref{setup}) depicts a sketch of the overall setup of the framework we adopt in this paper. 

In terms of Keplerian orbital elements, the lowest order expansion of the Hamiltonian that governs Mercury's secular evolution is \citep{MD99}
\begin{align}
\mathcal{H} &= - \frac{\mathcal{G} m m_2}{a_2} \bigg[ \feone e^2 + \fetwo e e_2 \cos (\varpi - \varpi_2) \nonumber \\
&+ \fione s^2 + \fitwo s s_2 \cos (\Omega - \Omega_2) \bigg],
\label{HLL}
\end{align}
where $m$ is mass, $a$ is semi-major axis, $e$ is eccentricity, $s = \sin(i/2)$, $i$ is inclination, $\varpi$ is longitude of perihelion, $\Omega$ is the longitude of ascending node, and $f$'s are constants that depend on the semi-major axis ratio ($a/a_2 < 1$) only. The functional forms of the constants can be looked up in the published literature (e.g. \citealt{LeVerrier1855, 1961mcm..book.....B, MD99}) and are given in the Appendix of this paper for reference. The subscripts denote the ordering of planets in the Solar System (e.g. $a_2$ is Venus' semi-major axis) while for simplicity, orbital elements without a subscript correspond to Mercury.

Keplerian orbital elements do not constitute a canonically conjugated set. Consequently, in order to employ the Hamiltonian framework for further progress, we introduce 
\Poincare\ action-angle variables defined as:
\begin{align}
\Lambda &= \mu \sqrt{\mathcal{G} (M_{\odot}+m) a}, \ \ \ \ \ \ \ \ \ \ \lambda = \mathcal{M} + \varpi, \nonumber \\
\Gamma &= \Lambda (1 - \sqrt{1-e^2}) \approx \Lambda \ e^2/2, \ \ \ \ \gamma = - \varpi, \nonumber \\
Z &= (1-\Gamma)(1 - \cos(i)) \approx 2 \Lambda \ s^2, \ \ \ \sigma = - \Omega,
\label{PoincareAAcoords}
\end{align}
where $M_{\odot}$ is the mass of the Sun, $\mu = m M_{\odot} / (M_{\odot}+m) \simeq m$ is the reduced mass, and $\mathcal{M}$ is mean anomaly. In terms of the \Poincare\ variables, the Hamiltonian (\ref{HLL}) reads:
\begin{align}
\mathcal{H} &= - \frac{\mathcal{G}^2 M m m_2^3}{\Lambda_2^2}\bigg[ 2 \feone \left(\frac{\Gamma}{\Lambda} \right) \nonumber \\
&+ 2 \fetwo \sqrt{\frac{\Gamma}{\Lambda} \frac{\Gamma_2}{\Lambda_2}} \cos (\gamma - \gamma_2) + \frac{\fione}{2} \left(\frac{Z}{\Lambda} \right)  \nonumber \\
&+ \frac{\fitwo}{2} \sqrt{\frac{Z}{\Lambda} \frac{Z_2}{\Lambda_2}} \cos (\sigma - \sigma_2) \bigg].
\label{HLLPoincare}
\end{align}

An assumption inherent to adopting the above Hamiltonian as an adequate dynamical model is that Mercury resides sufficiently far away from any low-order mean motion commensurabilities with the other planets, such that the associated resonant angles remain in rapid circulation. Under this assumption, averaging over the said angles renders the semi-major axis (and therefore the action $\Lambda$) a constant of motion \citep{Morbybook}. Consequently, Hamiltonian (\ref{HLLPoincare}) constitutes a non-autonomous (i.e. time-dependent) two degrees of freedom system.

Upon switching to canonical cartesian coordinates, defined as:
\begin{align}
x &= \sqrt{2 \Gamma} \cos(\gamma) \ \ \ \ \ y = \sqrt{2 \Gamma} \sin(\gamma) \nonumber, \\
w &= \sqrt{2 Z} \cos(\sigma) \ \ \ \ z = \sqrt{2 Z} \sin(\sigma), 
\label{definecartPoincare}
\end{align}
the Hamiltonian (\ref{HLLPoincare}) takes on the following form:
\begin{align}
\mathcal{H} &= - \frac{\mathcal{G}^2 M m m_2^3}{\Lambda_2^2}\bigg[ \feone \left( \frac{x^2 + y^2}{\Lambda} \right) \nonumber \\
&+ \fetwo \left( \frac{x x_2 + y y_2}{ \sqrt{\Lambda \Lambda_2} } \right) + \frac{\fione}{4} \left( \frac{w^2 + z^2}{\Lambda} \right) \nonumber \\
&+ \frac{\fitwo}{4}  \left( \frac{w w_2 + z z_2}{ \sqrt{\Lambda \Lambda_2} } \right) \bigg].
\label{HLLcart}
\end{align}

In order to compute Mercury's dynamical evolution, we must supply the Hamiltonian (\ref{HLLcart}) with a functional form for the perturbing variables. As a leading order approximation, we may assume that the characteristic rate of Mercury's chaotic diffusion substantially exceeds that of the other planets\footnote{In other words, Mercury's intrinsic Chirikov diffusion is more vigorous than the stochastic pumping of its orbit by extrinsic chaotic perturbations \citep{Yellowbook}.} and adopt a periodic secular solution for Venus. Specifically, we take the secular evolution of Venus' orbit to be given by a linear superposition of seven eigen-modes (see \citealt{Laskar90,MD99,Morbybook}), corresponding to a Lagrange-Laplace like solution:
\begin{align}
x_2 &= \sqrt{\Lambda_2} \sum_{j = 2}^{\hat{N}_e} \bar{e}_{2,j} \cos \left( g_j t + \beta_j  \right) \nonumber \\
y_2 &= - \sqrt{\Lambda_2} \sum_{j = 2}^{\hat{N}_e} \bar{e}_{2,j} \sin \left( g_j t + \beta_j  \right)  \nonumber \\
w_2 &= \sqrt{\Lambda_2} \sum_{j = 2}^{\hat{N}_s} \bar{s}_{2,j} \cos \left( f_j t + \theta_j  \right) \nonumber \\
z_2 &= -\sqrt{\Lambda_2} \sum_{j = 2}^{\hat{N}_s} \bar{s}_{2,j} \sin \left( f_j t + \theta_j  \right),
\label{Venussolution}
\end{align}
where $g$'s \& $f$'s are eigenfrequencies, $\bar{e}_{2,i}$'s \& $\bar{s}_{2,i}$'s are scaled eigenmode amplitudes, $\beta_i$'s \& $\theta_i$'s are phases, and $\hat{N}$'s represent the total number of eigenmodes in the decomposition. Note that we have purposefully dropped the first eigenmode. This filters out an unphysical self-resonance from the system (note further that assuming Mercury to be a test-particle yields only very limited corrections to the Lagrange-Laplace solution because of Mercury's almost negligible mass). 

Some variations of the Lagrange-Laplace solution exist in the literature (e.g. \citealt{WorkemBurkem,Bretagnon,Laskar90}), however the decompositions are sufficiently similar that for our purposes it does not matter exactly which solution we choose. For definitiveness, we shall adopt the solution of \citet{WorkemBurkem}, which is thoroughly documented in Ch.7 of \citet{MD99}. For reference, the dominant secular frequencies\footnote{The $g_1$ and $f_1$ modes, that primarily govern Mercury's secular dynamics are shown in green. The principal perturbing modes, namely $g_5$ and $f_2$, are highlighted in orange. Additionally, the auxiliary perturbing $g_2$ mode is depicted in blue.} of the Solar System are shown in Table (\ref{frequencies}).

\begin{table}[t]
\caption{Secular Eigenfrequencies of the Solar System ("/yr)}
\centering
\begin{tabular}{c c c c c c c c c c}
\hline\hline
$i$ & 1 & 2 & 3 & 4 & 5 & 6 & 7 & 8  \\ [0.5ex] % inserts table %heading
\hline
$g $ & \textcolor{OliveGreen}{5.46} & \textcolor{NavyBlue}{7.34} & 17.32 & 18.0 & \textcolor{YellowOrange}{4.29} & 27.77 & 2.72 & 0.63 \\
$f$ & \textcolor{OliveGreen}{-5.2} & \textcolor{YellowOrange}{-6.57} & -18.74 & -17.63 & 0.0 & -25.73& -2.9 & -0.67 \\[1ex]
\hline
\end{tabular}
\label{frequencies}
\end{table}

A peculiar feature of the periodic decomposition of the Solar System's secular dynamics is that the $g_1,g_2$ and $g_5$ modes (that dominate Mercury's, Venus's, and Jupiter's eccentricity variations respectively) as well as the $f_1$ and $f_2$ modes (that dominate Mercury's \& Venus's inclination variations) occupy the same frequency range to within $\sim1-2$ arcsec/year whereas the remainder of the relevant\footnote{The amplitudes associated with $g_7$, $g_8$, $f_{7}$ and $f_{8}$ modes are negligibly small and play an insignificant role in Mercury's dynamical evolution.} modes is separated from this group by $\sim10$ arcsec/year or more.

Upon direct substitution of equations (\ref{Venussolution}) into the Hamiltonian (\ref{HLLcart}), fourteen harmonics of the form $(\gamma + g_i t + \beta_i)$ and $(\sigma + f_i t + \theta_i)$ are generated. Although such a system is not easily tractable analytically, the grouping of the secular frequencies suggests that only a few of these harmonics are dynamically significant, and the rest can be ignored (i.e. averaged over). In particular, since $\dot{\gamma} \simeq - g_1$ and $\dot{\sigma} \simeq - f_1$, the beat frequencies of $(\gamma + g_5)$ and $(\sigma + f_2)$ are much smaller than the rest of the terms. 

Thus, following \citet{Boue} we drop all but two principal harmonics from the Hamiltonian:
\begin{align}
\mathcal{H} &= - \frac{\mathcal{G}^2 M m m_2^3}{\Lambda_2^2} \bigg[ 2 \feone \left(\frac{\Gamma}{\Lambda} \right) \nonumber \\
&+ \sqrt{2} \fetwo \bar{e}_{2,5} \sqrt{\frac{\Gamma}{\Lambda}} \cos (\gamma + g_5 t + \beta_5) + \frac{\feone}{2} \left(\frac{Z}{\Lambda} \right) \nonumber \\
&+ \frac{\fetwo}{2\sqrt{2}} \bar{s}_{2,2} \sqrt{\frac{Z}{\Lambda}} \cos (\sigma + f_2 t + \theta_2) \bigg].
\label{HLLreduced}
\end{align}
We note that in addition to these harmonics, there exists an additional mode (namely the $g_2$ mode), that plays an important role in Mercury's chaotic evolution \citep{LithWu11}. We shall add this term and analyze its effects later. For now, we continue with the simplified Hamiltonian (\ref{HLLreduced}).

The explicit time dependence in Hamiltonian (\ref{HLLreduced}) can be eliminated by extending the phase space \citep{Morbybook} to accommodate an additional action, $\mathbb{T}$ conjugate to $t$ (temporarily, this Hamiltonian will be characterized by 3 degrees of freedom):
\begin{align}
\mathcal{H} &= - \frac{\mathcal{G}^2 M m m_2^3}{\Lambda_2^2} \bigg[ 2 \feone \left(\frac{\Gamma}{\Lambda} \right) \nonumber \\ 
&+ \sqrt{2} \fetwo \bar{e}_{2,5} \sqrt{\frac{\Gamma}{\Lambda}} \cos (\gamma + g_5 t + \beta_5) \frac{\feone}{2} \left(\frac{Z}{\Lambda} \right) \nonumber \\ 
&+ \frac{\fetwo}{2 \sqrt{2}} \bar{s}_{2,2} \sqrt{\frac{Z}{\Lambda}} \cos (\sigma + f_2 t + \theta_2) \bigg] + \mathbb{T}.
\label{HLLreducedT}
\end{align}

\begin{figure*}[t]
\centering
\includegraphics[width=0.8\textwidth]{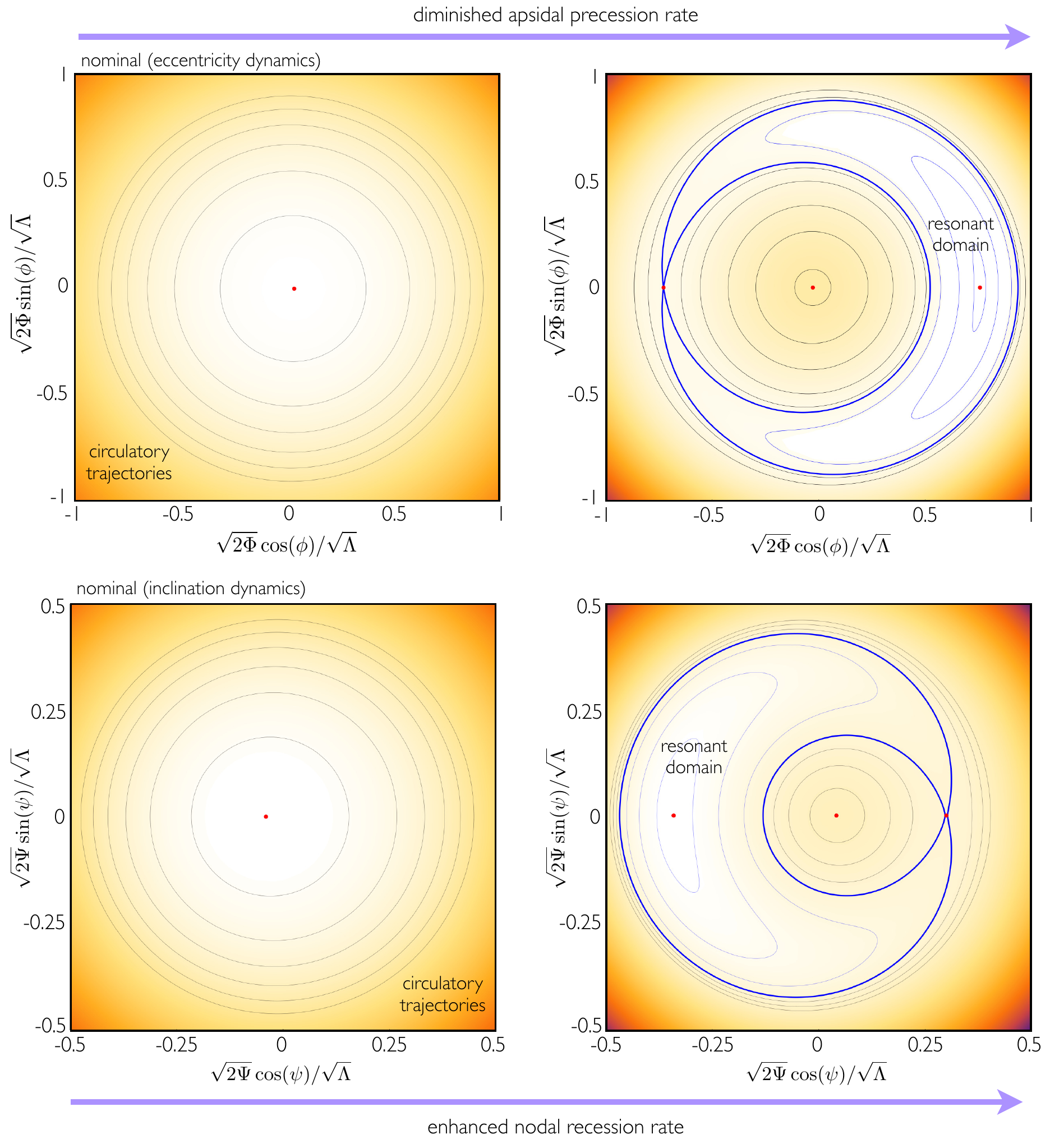}
\caption{Phase-space portraits of the individual degrees of freedom of Hamiltonian (\ref{Hsfmr}). The level curves of $\mathcal{H}$ are shown in terms of global cartesian coordinates, scaled such that in the vicinity of the origin, the radial distance is approximately $e$ in the top panels and $i$ in the bottom panels. The left panels are plotted at nominal Solar System parameters and show harmonic oscillator like dynamics, where the phase-space is foliated in elliptical orbits nested around slightly off-center stable fixed points. Conversely, the panels on the right depict pendulum-like dynamics that correspond to a system where the natural apsidal precession rate of Mercury has been reduced while its nodal recession rate has been enhanced manually. The modified version of the system characterizes Mercury's dynamical state at the onset of large-scale instability. Accordingly, the existence of homoclinic curves (shown in blue) as well as the associated resonant trajectories is readily evident in these panels.}
\label{PSHSFMR}
\end{figure*}

Let us now perform a canonical transformation of variables that arises from the following generating function of the second kind:
\begin{equation}
\tilde{\mathbb{G}}_2 = (\gamma + g_5 t + \beta_5) \Phi + (\sigma + f_2 t + \theta_2) \Psi + t \Xi.
\end{equation}
An application of the transformation equations yields new action-angle variables:
\begin{align}
\label{transformone}
\Phi &= \Gamma \ \ \ \ \ \ \ \ \ \phi = \gamma + g_5 t + \beta_5, \nonumber \\
\Psi &= Z \ \ \ \ \ \ \ \ \ \psi = \sigma + f_2 t + \theta_2,\nonumber  \\
\Xi &= \mathbb{T} - g_5 \Phi - f_2 \Psi \ \ \ \ \ \ \xi = t.
\end{align}
Accordingly, the Hamiltonian takes the form:
\begin{align}
\mathcal{H} &= \Xi - \frac{\mathcal{G}^2 M m m_2^3}{\Lambda_2^2} \bigg[ 2 \feone \left(\frac{\Phi}{\Lambda} \right) \nonumber \\
&+ \sqrt{2} \fetwo \bar{e}_{2,5} \sqrt{\frac{\Phi}{\Lambda}} \cos (\phi) + \frac{\feone}{2} \left(\frac{\Psi}{\Lambda} \right) \nonumber \\ 
&+ \frac{\fetwo}{2 \sqrt{2}} \bar{s}_{2,2} \sqrt{\frac{\Psi}{\Lambda}} \cos (\psi) \bigg] + g_5 \Phi + f_2 \Psi.
\label{HLLphipsi}
\end{align}
Because $\partial \mathcal{H}/ \partial \xi = 0$, $\Xi$ is a constant of motion and can therefore be dropped from the Hamiltonian. 

As briefly alluded to in the introduction of this paper, in addition to planet-planet interactions, it has been shown that the apsidal precession arising from general relativistic effects plays an important, stabilizing role in Mercury's secular dynamics \citep{Laskar08,BatLaugh,Laskar09}. To leading order in $e$, this precession can be accounted for by adding a term proportional to $\Phi$ (equivalently, $\Gamma$) to $\mathcal{H}$:
\begin{align}
\mathcal{H} &= - \frac{\mathcal{G}^2 M m m_2^3}{\Lambda_2^2} \bigg[ 2 \feone \left(\frac{\Phi}{\Lambda} \right) \nonumber \\
&+ \sqrt{2} \fetwo \bar{e}_{2,5} \sqrt{\frac{\Phi}{\Lambda}} \cos (\phi) + \frac{\feone}{2} \left(\frac{\Psi}{\Lambda} \right) \nonumber \\
&+ \frac{\fetwo}{2 \sqrt{2}} \bar{s}_{2,2} \sqrt{\frac{\Psi}{\Lambda}} \cos (\psi) \bigg] \nonumber \\
&+ \left(g_5 - \frac{3 \mathcal{G} M n}{a c^2} \right) \Phi + f_2 \Psi,
\label{HLLgr}
\end{align}
where $c$ is the speed of light.

Recall that the Hamiltonian (\ref{HLLgr}) governs the secular three-body problem. Let us now extend the above Hamiltonian to account for interactions between Mercury and all Solar System planets. As before, we shall only retain the $g_5$ and $f_2$ modes in each planet's assumed secular solution. This allows us to simply introduce six additional clones of the bracketed expression in the Hamiltonian (\ref{HLLgr}) and sum over them. The Hamiltonian thus takes on the following form:
\begin{align}
\mathcal{H} &= \left(F_{\rm{GR}} + \Feone + g_5 \right) \Phi + \Fetwofive \sqrt{\Phi} \cos(\phi) \nonumber \\
&+ \left(\Fione + f_2 \right) \Psi + \Fitwo \sqrt{\Psi} \cos(\psi).
\label{HLLgrF}
\end{align}
For reference, the coefficients read:
\begin{align}
\Feone &= -\sum_{j = 2}^8 \left( \frac{\mathcal{G} m m_j}{a_j} \frac{2}{\Lambda} \feonej \right) = -2.75 \times 10^{-5}  \nonumber \\
\Fetwofive &= -\sum_{j = 2}^8 \left( \frac{\mathcal{G} m m_j}{a_j} \sqrt{\frac{2}{\Lambda}} \fetwoj \bar{e}_{j,5} \right) = 2.77 \times 10^{-10} \nonumber \\
\Fione &= -\sum_{j = 2}^8 \left( \frac{\mathcal{G} m m_j}{a_j} \frac{1}{2 \Lambda} \fionej \right) = 2.75 \times 10^{-5} \nonumber \\
\Fitwo &= -\sum_{j = 2}^8 \left( \frac{\mathcal{G} m m_j}{a_j} \sqrt{\frac{1}{8 \Lambda}} \fitwoj \bar{s}_{j,2} \right) = - 2.01 \times 10^{-10}\nonumber \\
F_{\rm{GR}} &= - \left( \frac{3 \mathcal{G} M n}{a c^2} \right)  = -1.99  \times 10^{-6}.
\label{coeffsF}
\end{align}
The numerical values of the coefficients are given in units of $M_{\odot}$, AU and years, such that $G = 4 \pi^2$.

The Hamiltonian (\ref{HLLgrF}) is a trivial canonical translation away from that of a pair of decoupled simple harmonic oscillators. Consequently, the phase-space portrait in either degree of freedom is a family of circles, nested around an elliptical fixed point that resides on the cartesian $x$-axis (i.e. corresponding to $\phi = 0$; $\psi = \pi$). 

\subsection{A Nonlinear Integrable Model}

Within the framework of the Hamilonian (\ref{HLLgrF}), resonance is ill-defined because homoclinic curves are absent from phase space. Thus, in order to properly define secular resonances, we must introduce nonlinearity into $\mathcal{H}$. The relevant terms in the series expansion of the gravitational potential are those proportional to $\propto e^4$ and $\propto s^4$ \citep{Sid1990,MD99}. Accordingly, following the transformation (\ref{transformone}), the Hamiltonian reads:
\begin{align}
\mathcal{H} &= \left(F_{\rm{GR}} + \Feone + g_5 \right) \Phi + \Fethree \Phi^2 \nonumber \\
&+ \Fetwofive \sqrt{\Phi} \cos(\phi) + \left(\Fione + f_2 \right) \Psi \nonumber \\
&+ \Fithree \Psi^2 + \Fitwo \sqrt{\Psi} \cos(\psi).
\label{Hsfmr}
\end{align}
The newly introduced constants are\footnote{Note that if one chooses to adopt the exact form of \Poincare\ action-angle coordinates (\ref{PoincareAAcoords}), non-linear action terms in $\mathcal{H}$ arise even at order $e^2$ and $i^2$. If desired, the corresponding contributions in Hamiltonian (\ref{Hsfmr}) can then be retained in $\Fethree$ and $\Fithree$.}:
\begin{align}
\Fethree &= -\sum_{j = 2}^8 \left( \frac{\mathcal{G} m m_j}{a_j} \left( \frac{2}{\Lambda} \right)^2 \fethreej \right) = -24.31 \nonumber \\
\Fithree &= -\sum_{j = 2}^8 \left( \frac{\mathcal{G} m m_j}{a_j} \left( \frac{1}{2 \Lambda} \right)^2 \fithreej \right) = -64.34.
\label{coeffsF3}
\end{align}

Individually, the two degrees of freedom are described by pendulum-like Hamiltonians, possessing D'Almbert characteristics \citep{1982amdc.proc..153H}. Hamiltonians of this sort appear in various aspects of celestial mechanics \citep{Wisdom83, 1986CeMec..38..175W, 1990CeMDA..47...99H, Sid1990, 2000Icar..148..282N, 2001Icar..150..104N, Morbybook, 2013A&A...556A..28B, 2013ApJ...774..129D}, as well as other dynamical systems such as high-intensity particle accelerator beams \citep{1994PhRvL..73.1247G,2010NIMPA.618...37B}. Owing to their wide-spread applicability, Hamiltonians of the form (\ref{Hsfmr}) are generally referred to as second fundamental models for resonance \citep{1983CeMec..30..197H}.

Because each degree of freedom is separately integrable, chaotic motion cannot arise within the framework of Hamiltonian (\ref{Hsfmr}). However, this Hamiltonian can still be used to qualitatively understand the conditions under which nonlinear secular resonances will give rise to instability.

Phase-space portraits of both degrees of freedom of the Hamiltonian (\ref{Hsfmr}) are shown in Figure (\ref{PSHSFMR}). With nominal parameters (panels on the left), the phase space portraits are quite reminiscent of the linear model (\ref{HLLgrF}). That is, phase-space is foliated in ellipses surrounding a stable equilibrium point. However, the situation is markedly different if slightly different parameters are chosen. 

\begin{figure*}[t]
\centering
\includegraphics[width=1\textwidth]{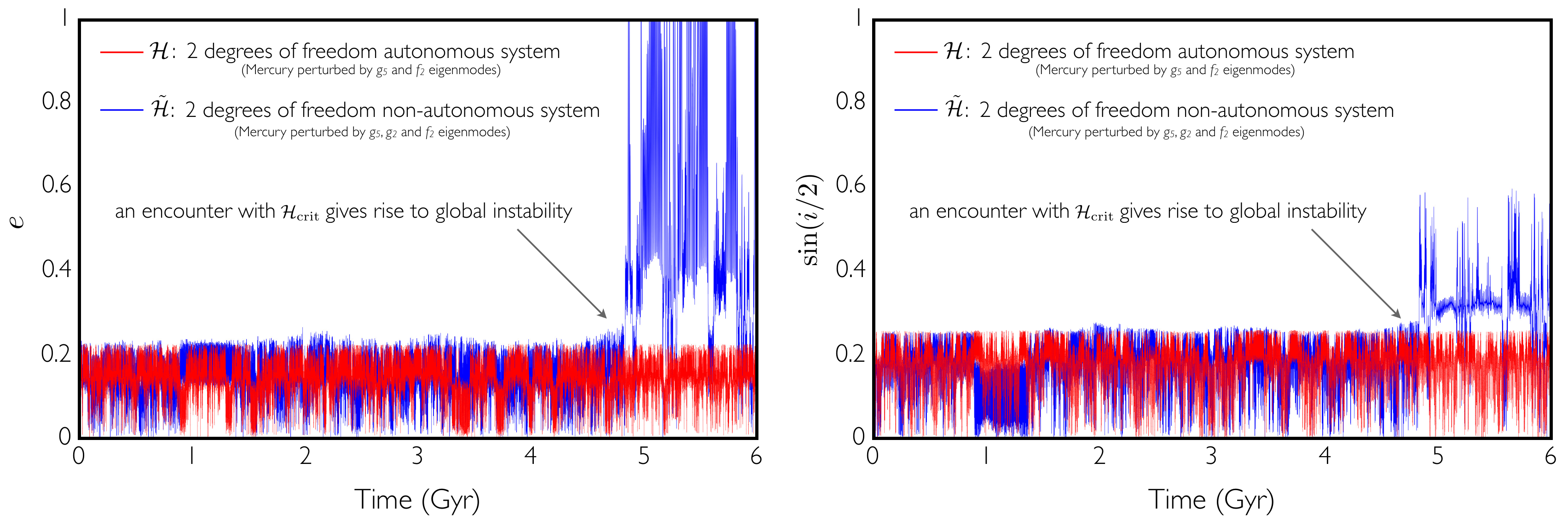}
\caption{Long-term dynamical evolution of Mercury. The orbital solutions were obtained by numerically integrating the equations of motion stemming from the autonomous 2 degree of freedom Hamiltonian (\ref{HtwoDOF}) (shown in red) and the non-autonomous 2 degree of freedom Hamiltonian (\ref{HthreeDOF}) (shown in blue). As Mercury's initial condition for the integrations, the phases and amplitudes of the $g_1$ and $f_1$ eigenmodes were adopted.  Evidently, the autonomous system adequately represents the stochastic properties of Mercury's evolution on multi-Myr timescales. However, transitions from bounded to unbounded chaos are only captured by a more complex, non-autonomous system. As discussed in the text, the onset of large-scale instability is facilitated by the system's acquisition of a critical value of $\tilde{\mathcal{H}}$ (see also Figure \ref{Hoft}) and additionally corresponds to the system's entrance into the $\nu_5$ and $\nu_{12}$ secular resonances.}
\label{ei}
\end{figure*}

To begin with, note that the system is rather close to exact secular commensurability. That is, constants that multiply the linear action terms in $\mathcal{H}$ are close to zero: $(F_{\rm{GR}} + \Feone + g_5 )/g_5 \simeq - 0.42$; $ (\Fione + f_2 )/f_2 \simeq 0.14$. Now suppose we introduce additional forcing that slows down Mercury's apsidal precession rate and speeds up its nodal recession rate. For the sake of argument, let us further assume that the modulation is such that the signs (but not magnitudes) of the constants in front of the linear action terms change. 

The corresponding phase-space portraits are shown in the right panels of Figure (\ref{PSHSFMR}). Evidently, the aforementioned manual modulation gives rise to separatrixes in both degrees of freedom and associated resonant trajectories (croissant-shaped curves) appear. The equilibria shown in the left panels of Figure (\ref{PSHSFMR}) correspond to the resonant equilibria in the right panels, which in turn reside at high eccentricity and inclination (see also \citealt{Boue}). Consequently, a modulation of the linear terms of the Hamiltonian can carry the trajectory to a part of phase space characterized by a sufficiently high eccentricity\footnote{It should be understood that Mercury's acquisition of high eccentricity and inclination is not an adiabatic process \citep{1984PriMM..48..197N} which allows it to remain at an equilibrium point as system parameters slowly change. Still, this exercise remains useful as an illustrative example.} to permit close encounters between Mercury and Venus \citep{BatLaugh} as well as Mars and the Earth \citep{Laskar09}.

\subsection{A Chaotic Model With 2 Degrees of Freedom}

The manual modulation of the secular frequencies invoked above in fact arises naturally from action-coupling between the two degrees of freedom \citep{LithWu11,Boue}. In particular, the relevant fourth-order term that governs this coupling is of the form $\propto e^2 s^2$. Upon incorporation of this term, the Hamiltonian takes on the following form:
\begin{empheq}[box=\fbox]{align}
\mathcal{H} &= \left(F_{\rm{GR}} + \Feone + g_5 \right) \Phi + \Fethree \Phi^2  \nonumber \\
&+ \Fetwofive \sqrt{\Phi} \cos(\phi) + \left(\Fione + f_2 \right) \Psi + \Fithree \Psi^2 \nonumber \\ 
&+ \Fitwo \sqrt{\Psi} \cos(\psi) + \Fei \Phi \Psi,
\label{HtwoDOF}
\end{empheq}
where the associated constant reads:
\begin{equation}
\Fei= -\sum_{j = 2}^8 \left( \frac{\mathcal{G} m m_j}{a_j} \left( \frac{1}{\Lambda} \right)^2 \feij \right) = 3 \times 10^2.
\label{Fei}
\end{equation}

The final term in Hamiltonian (\ref{HtwoDOF}) breaks its integrability and allows for the possibility of stochastic evolution. Accordingly, this gives rise to the quasi-random alteration of the secular frequencies.

Numerical integration of the equations of motion derived from the Hamiltonian (\ref{HtwoDOF}) reveals chaotic eccentricity and inclination dynamics over a broad parameter range. A particular realization of the long-term stochastic evolution of Mercury is shown in Figure (\ref{ei}) with red curves, where the adopted initial conditions correspond to the amplitudes and phases of the $g_1$ and $f_1$ eigen-modes of the Lagrange-Laplace solution\footnote{Quantitatively, the adopted initial conditions are close to Mercury's present orbital state and choosing the latter does not alter the results in any meaningful way. However the use of the $g_1$ and $f_1$ eigenmode amplitudes and phases as initial conditions is formally more appropriate, since the majority of contributing secular modes have been averaged over in equation (\ref{HLLreduced}).}. 

Upon linearization of the equations of motion and an application of the MEGNO algorithm \citep{2000A&AS..147..205C} to the system at hand, we obtain a numerical estimate for the Lyapunov time of $\tau_{\rm{L}} = 1.1$ Myr. This value is in satisfactory agreement with that obtained using more complex perturbative and N-body methods \citep{Laskar89,Quinn1991,SussmanWisdom1992,BatLaugh}. Consequently, it seems likely that on timescales comparable to $\tau_{\rm{L}}$, the simplified system described by Hamiltonian (\ref{HtwoDOF}) captures the chaotic properties of Mercury's actual orbit in an adequate manner.

\section{Chaotic Diffusion of Eccentricity and Inclination}

With a simple model at hand, let us now explore the chaotic properties of Mercury's secular evolution. Particularly, in this section we shall analytically derive Mercury's Lyapunov time as well as its chaotic diffusion coefficient related to $e$ and $i$. As a first step of this calculation, it is worthwhile to delineate the admissible region of phase-space on which Mercury's secular dynamics reside.

\subsection{The Admissible Domain of $\mathcal{H}$}

The flow governed by Hamiltonian (\ref{HtwoDOF}) is constrained by the autonomous nature (i.e. the conservation) of $\mathcal{H}$. That is, despite chaotic diffusion, there are forbidden regions of phase space that the system can not explore. An implicit assumption inherent to this assertion is that at fixed values of the angles, $\Phi$ is a decreasing function of $\Psi$ (and vice-versa). Mathematically, the admissible domain is characterized by null imaginary components of the actions. Consequently, provided a value of $\mathcal{H}$ corresponding to the initial conditions (let us denote it $\mathcal{H}_0$), the bounding curves are obtained by sequentially equating $\Phi$ and $\Psi$ to zero:
\begin{align}
\mathcal{H}_0 &= \left(F_{\rm{GR}} + \Feone + g_5 \right) \Phi + \Fethree \Phi^2 
+ \Fetwofive \sqrt{\Phi} \cos(\phi) \nonumber \\
\mathcal{H}_0 &= \left(\Fione + f_2 \right) \Psi + \Fithree \Psi^2 + \Fitwo \sqrt{\Psi} \cos(\psi).
\label{boundaries}
\end{align}

%The expressions (\ref{boundaries}) further allow one to compute the maximal values of $\Phi_{\rm{max}}$ and $\Psi_{\rm{max}}$ (and therefore $e_{\rm{max}}$ and $s_{\rm{max}}$) that can be attained by chaotic diffusion for a given set of initial conditions\footnote{For the case shown in Figure (\ref{ei}), the expressions (\ref{boundaries}) yield $e_{\rm{max}} = 0.21$ and $s_{\rm{max}} = 0.26$ in agreement with the simulations.}. 

The fact that with nominal parameters the admissible region does not extend to very high eccentricity is of great importance for the stability of the inner Solar System, as it ensures that (to the extent that the Hamiltonian (\ref{HtwoDOF}) is a good approximation to the real dynamics) chaotic diffusion remains confined. Such behavior can be readily observed in Figure (\ref{ei}), where the eccentricity and inclination evolution (governed by Hamiltonian \ref{HtwoDOF} - red curves) stemming from numerical integration of equations of motion (see Appendix) appears perfectly bounded. 

\subsection{High-Order Secular Resonances}

\begin{figure*}[t]
\includegraphics[width=1\textwidth]{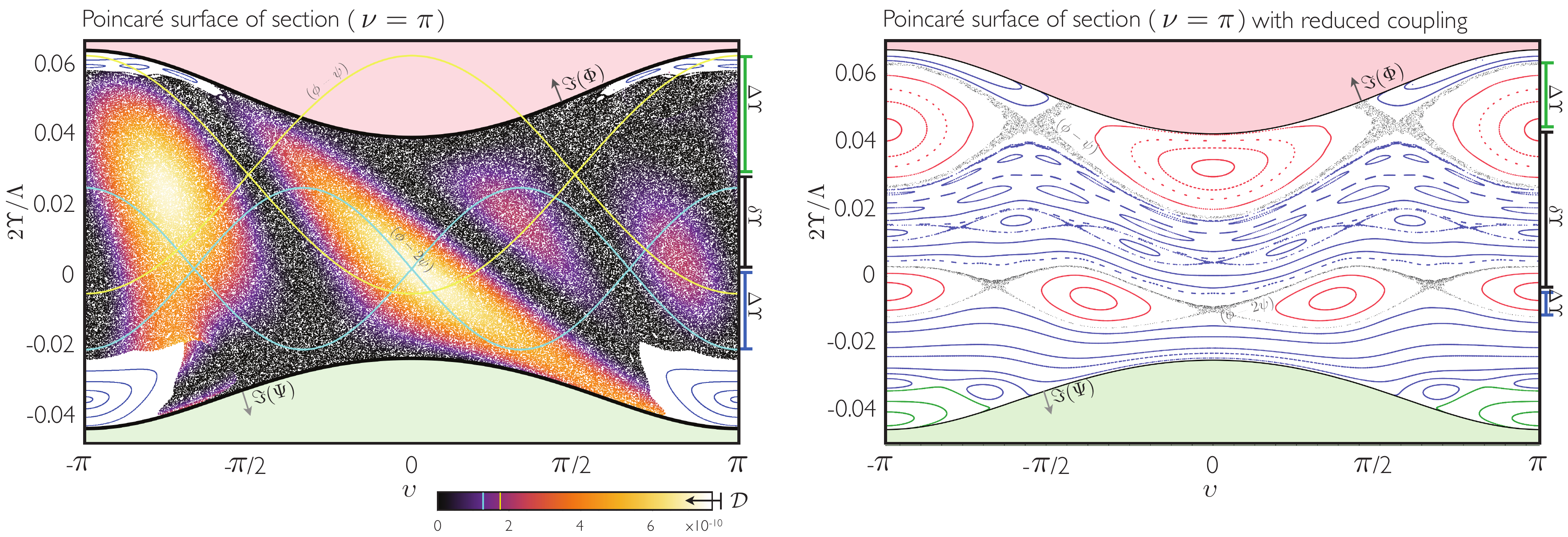}
\caption{\Poincare\ surfaces of section of the 2 degree of freedom autonomous Hamiltonian, $\mathcal{H}$ (expression \ref{HtwoDOF}). The left panel depicts a surface of section with nominal parameters. Specifically, the black-gold points depict a numerically obtained surface of section and the color represents the local chaotic diffusion coefficient, computed as the square of the change in action between sequential section points divided by the corresponding change in time. Blue curves denote quasi-periodic trajectories. The critical curves (in the pendulum approximation) of the $(\phi-\psi)$ resonance (gold), and $(\phi-2\psi)$ resonance (cyan) are over-plotted on the section. The resonance widths, $\Delta\Upsilon$ as well as the distance between the resonances, $\delta\Upsilon$ are depicted on the side of the panel. The curves bounding the admissible domain within which the actions are real are additionally labeled. The $y-$axis of the section is scaled such that $2\Upsilon/\Lambda\simeq e^2-i^2$. The right panel depicts an equivalent surface of section, but with a coupling parameter ($\Fei$) that has been reduced by a factor of $0.6$. Naturally, as the reduction in the nonlinear coupling brings the system closer to integrability, the majority of the phase-space on the right panel is occupied by quasi-periodic trajectories. The vicinity of the unperturbed separatrixes of the analytically identified $(\phi-\psi)$ and $(\phi-2\psi)$ resonances are encompassed by thin chaotic layers and are shown with black points. Meanwhile, the corresponding quasi-periodic resonant trajectories are shown in red. Note that in addition to the two primary resonances, there also exists an intricate web of yet higher order secular resonances. Although the angles associated with these resonances also undergo chaotic evolution under nominal parameters, their contribution to Mercury's stochasticity is sub-dominant.}
\label{Psection}
\end{figure*}

Let us now identify the dominant features of the dynamical structure within the admissible region of phase space. To begin with, recall that the principal harmonics $\phi$ and $\psi$ present in $\mathcal{H}$ are in circulation given nominal parameters, and homoclinic curves are noticeably absent from the phase-space portraits (see Figure \ref{PSHSFMR}). However, the lack of overlap of resonances associated with $\phi$ and $\psi$ clearly does not imply a lack of chaos, and indeed we must look to higher-order resonances to explain the stochastic behavior exhibited by Mercury. We shall do so by making use of canonical perturbation theory.

There exist numerous flavors of canonical perturbation theory \citep{Goldstein,Yellowbook}, each in principle as appropriate as the next, depending on the problem at hand. In this work, we wish to retain the ability to carry out the perturbation series to order higher than two. In practice, this is best accomplished by employing Lie transformation methods \citep{1966PASJ...18..287H, 1969CeMec...1...12D, Morbybook} and this is the approach we adopt here.

We begin by separating Hamiltonian (\ref{HtwoDOF}) into a trivially integrable component:
\begin{align}
\check{\mathcal{H}} &= \left(F_{\rm{GR}} + \Feone + g_5 \right) \Phi + \Fethree \Phi^2  \nonumber \\
&+\left(\Fione + f_2 \right) \Psi + \Fithree \Psi^2 + \Fei \Phi \Psi,
\label{H0}
\end{align}
and a perturbation:
\begin{align}
\mathcal{H}' &= \epsilon\bigg[\Fetwofive \sqrt{\Phi} \cos(\phi) + \Fitwo \sqrt{\Psi} \cos(\psi)\bigg],
\label{H1}
\end{align}
where $\epsilon$ is a formal ``label" of order, which we will set to unity later. With the above expressions, the homologic equation:
\begin{equation}
\mathcal{H}' + \left\{\check{\mathcal{H}}, \chi \right\} = 0,
\label{hologic}
\end{equation}
where $\{\}$ is the Poisson bracket, is satisfied by the generating Hamiltonian
\begin{eqnarray}
\chi &=& \epsilon\bigg[\frac{\Fetwofive \sqrt{\Phi}}{F_{\rm{GR}} + \Feone + g_5+2\Fethree \Phi +\Fei \Psi} \sin(\phi) \nonumber \\
&+& \frac{\Fitwo \sqrt{\Psi}}{\Fione + f_2+2\Fithree \Psi +\Fei \Phi} \sin(\psi)\bigg].
\label{chi}
\end{eqnarray}
Written out explicitly to third order in $\epsilon$, the averaged Hamiltonian takes the form \citep{Morbybook}:
\begin{align}
\bar{\mathcal{H}} &= \check{\mathcal{H}} + \{\mathcal{H}',\chi\} +\frac{1}{2} \{\{\check{\mathcal{H}},\chi\},\chi\} + \frac{1}{2} \{\{\mathcal{H}',\chi\},\chi\}  \nonumber \\
&+ \frac{1}{6} \{\{\{\check{\mathcal{H}},\chi\},\chi\}, \chi\} + \mathcal{O}(\epsilon^4).
\label{Have}
\end{align}

As expliciated by equation (\ref{Have}), the averaging procedure (i.e. the application of the Lie transform under the flow of $\chi$), eliminates harmonics of order $\epsilon$ (i.e. equation \ref{hologic}) at the expense of introducing new harmonics at orders $\epsilon^2, \epsilon^3, ...\ $ (see \citealt{1993CeMDA..55..101M,1993CeMDA..55..131M} for an in-depth discussion). Specifically, at order $\epsilon^2$ the averaging process generates the angles $(2\phi), (2\psi), (\phi-\psi)$, and $(\phi + \psi)$, while at order $\epsilon^3$ the procedure additionally yields $(\phi), (\psi), (3\phi), (3\psi), (2\phi-\psi), (2\phi+\psi), (\phi-2\psi)$, and $(\phi + 2\psi)$. Naturally, as $\phi$ and $\psi$ individually obey D'Almbert rules, so do their combinations.

Following \citet{Chirikov1979}, we analyze the generated harmonics independently, and examine the equilibria of the associated resonances. Since the leading order perturbation is now proportional to $\epsilon^2$, let us perform a canonical transformation of variables such that the new angles correspond to the novel second-order harmonics introduced by the averaging process. This can be accomplished by employing the following generating function of the second type:
\begin{align}
\mathbb{G}_2 = ( \phi-\psi )\Upsilon/2 + \left( \phi+\psi \right) \mathcal{V}/2,
\label{Gtwo}
\end{align}
which yields the action-angle coordinates:
\begin{align}
\upsilon = (\phi-\psi)/2, \ \ \ \ \ \    \Upsilon = \Phi - \Psi, \nonumber \\
\nu = (\phi+\psi)/2, \ \ \ \ \ \ \mathcal{V} = \Phi + \Psi.
\end{align}

Given that both $\phi$ and $\psi$ circulate rapidly into the same (negative) direction, so does the newly defined angle $\nu$. However, a cursory inspection of the Hamiltonian (provided nominal parameters) reveals that although neither $\phi$ nor $\psi$ undergo bounded oscillations, their time-derivatives are nearly identical, $\dot{\phi} \approx \dot{\psi}$, meaning that the beat angle $\upsilon$ can be expected to resonate\footnote{In fact, $\upsilon$ is a chaotic angle, as was first shown by \citet{Laskar89} (see also \citealt{SussmanWisdom1992,LithWu11}).}. With this consideration in mind, we construct a \Poincare\ surface of section with respect to the rapidly circulating angle at $\nu = \pi$. 

The surface of section for the nominal value of the Hamiltonian $\mathcal{H}_0$ is shown in the left panel of Figure (\ref{Psection}). The pink and green regions in the Figure show the inadmissible part of phase space, while thick black curves denote the boundaries of the admissible region (equations \ref{boundaries}). Quasi-periodic trajectories are shown as blue curves, while the chaotic sea is depicted with black-gold points. The color scale inherent to the chaotic sea represents the local chaotic diffusion coefficient $\mathcal{D}_{\Upsilon}$, which is computed as the square of the change in action $\Upsilon$ divided by the time difference between successive section points. Note that the local diffusion coefficient appears to track the deformed structure of the underlying resonances\footnote{See \citet{1985PhRvA..32.2413M} for an in-depth discussion of the non-uniformity of diffusion in a chaotic layer.}.

Upon examination, it is immediately clear that the vast majority of phase space is occupied by chaotic trajectories, signaling gross overlap of at least two high-order secular resonances \citep{Chirikov1959}. In other words, there exist at least two high-order secular resonances whose equilibria lie within the admissible region. In an effort to identify the overlapping resonances, let us begin by plotting the critical curve of the $(\phi-\psi)$ resonance on the surface of section (\ref{Psection}).

The coefficient of the harmonic is obtained at order $\epsilon^2$ in the perturbation series:
\begin{align}
\mathcal{C}_{\epsilon^2}&\cos(\phi - \psi) = \bigg( \left\{\mathcal{H',\chi} \right\} + \frac{\left\{ \left\{ \check{\mathcal{H}},\chi \right\} , \chi \right\}  }{2} \bigg)_{(\phi-\psi)} \nonumber \\
&= \big( \Fetwofive \Fitwo \Fei \sqrt{\Phi} \sqrt{\Psi} \big) \cos(\phi - \psi)  \nonumber \\
&\times \bigg[ \big( 4 (F_{\rm{GR}} + \Feone + g_5 + 2 \Fethree \Phi + \Fei \Psi)^{2} \big)^{-1} \nonumber \\
&+ \big(4 (\Fione + f_2 + 2 \Fithree \Psi + \Fei \Phi )^2 \big)^{-1} \bigg].
\label{phipsi}
\end{align}
Obviously, the function $\mathcal{C}$ exhibits a rather complex dependence on the actions. With the goal of intelligibility in mind, here we replace $\Phi$ and $\Psi$ in equation (\ref{phipsi}) with their nominal values, corresponding to the amplitudes of the $g_1$ and $f_1$ eigenmodes of Mercury's Lagrange-Laplace decomposition. As such, the amplitude of the harmonic is assumed to be constant.

Augmenting the integrable Hamiltonian (\ref{H0}) with an exclusive perturbation (\ref{phipsi}), the Hamiltonian is cast into a familiar pendulum-like form:
\begin{align}
\mathcal{H}_{\epsilon^n} = \mathcal{F}_{\epsilon^n} (\Upsilon - \Upsilon_0^{\epsilon^n})^2 + \mathcal{C}_{\epsilon^n} \cos (n \upsilon),
\label{pendulum}
\end{align}
where $\Upsilon_0^{\epsilon^n}$ is an action corresponding to the fixed points of $\mathcal{H}_{\epsilon^n}$ and $n$ is the order of resonance. Completing the square, the parameters of $\mathcal{H}_{\epsilon^2}$ take on the following forms:
\begin{align}
\mathcal{F}_{\epsilon^2} &= (\Fethree+\Fithree-\Fei)/4, \nonumber \\
\Upsilon_0^{\epsilon^2} &= (F_{\rm{GR}} + \Feone + g_5-\Fione - f_2+\Fethree \mathcal{V} \nonumber \\
&- \Fithree \mathcal{V})(\Fethree-\Fei+\Fithree)^{-1},
\end{align}
where $\mathcal{V}$ is evaluated at nominal actions.

The separatrix of the corresponding resonance is shown as a gold curve in action-angle variables in the left panel of Figure (\ref{Psection}). By deforming the dynamics of the  resonance into that a pendulum (by forcing $\mathcal{C}$ to be constant), we allow the critical curve to not be constrained by the admissible domain of the Hamiltonian (\ref{HtwoDOF}). Although a more careful treatment of perturbation series can remedy this inconsistency, at our desired level of approximation this does not constitute a significant drawback.

Because the remaining second-order harmonics $(2\phi)$, $(2\psi)$ and $(\phi + \psi)$ circulate rapidly, their equilibria lie well outside of the admissible region. Accordingly, these harmonics do not contribute to the stochasticity of the evolution. Therefore, the $(\phi-\psi)$ resonance is overlapped by a harmonic of order higher than two. 

Analyzing the angles proportional to $\epsilon^3$ in the same way as above, we find the equilibria of all resonances except $(\phi - 2\psi)$ to reside outside of the admissible region. The amplitude of the $(\phi - 2\psi)$ resonance is computed in the following way:
\begin{align}
\mathcal{C}_{\epsilon^3} &\cos(\phi - 2\psi) = \bigg(\frac{ \left\{ \left\{ \mathcal{H}', \chi \right\}, \chi \right\}}{2} + \frac{ \left\{ \left\{ \left\{ \check{\mathcal{H}},\chi \right\} , \chi \right\}, \chi \right\}  }{6} \bigg)_{(\phi-2\psi)} \nonumber \\
&=\big( \Fetwofive^2 \Fitwo \Fei \Phi \sqrt{\Psi} \big) \cos(\phi-2\psi) \bigg[ 2 \Fei (F_{\rm{GR}} + \Feone  \nonumber \\ 
&+ g_5 + 2 \Fethree \Phi + \Fei \Psi)^3(\Fione + f_2 + \Fei \Phi + 2 \Fithree \Psi)^{-3}  \nonumber \\
& -6 \Fethree(F_{\rm{GR}} + \Feone + g_5 + 2 \Fethree \Phi + \Fei \Psi)^2(\Fione \nonumber \\
&+ f_2 + \Fei \Phi + 2 \Fithree \Psi)^{-2} + \Fei (F_{\rm{GR}} + \Feone + g_5 \nonumber \\
&+ 2 \Fethree \Phi + \Fei \Psi)^2 (\Fione + f_2 + \Fei \Phi + 2 \Fithree \Psi)^{-2} \nonumber \\
&- 4 \Fethree( F_{\rm{GR}} + \Feone + g_5 +2 \Fethree \Phi  + \Fei \Psi) \nonumber \\
&\times (\Fione + f_2 + \Fei \Phi + 2 \Fithree \Psi)^{-1} + 2 \Fethree + \Fei \bigg]\nonumber \\
&\times \big(12 (F_{\rm{GR}} + \Feone + g_5 + 2 \Fetwofive \Phi + \Fei \Psi)^4 \big)^{-1}.
\label{phi2psi}
\end{align}

The perturbing $n=3$, $(\phi-2\psi)$ resonance can be molded into the form (\ref{pendulum}) through a variable change arising from the generating function
\begin{align}
\bar{\mathbb{G}}_2 = (\phi - 2\psi)\bar{\Upsilon}/3 + \left( \phi + \psi \right) \bar{\mathcal{V}}/2.
\label{Gtwobar}
\end{align}
The new variables are related to the old ones through:
\begin{align}
\bar{\upsilon} &= (\phi-2\psi)/3, \ \ \ \ \ \ \bar{\Upsilon} = \Phi - \Psi, \nonumber \\
\bar{\nu} &= (\phi+\psi)/2, \ \ \ \ \ \ \ \ \bar{\mathcal{V}} = 2 (2\Phi + \Psi)/3.
\end{align}
Accordingly, the constants of the Hamiltonian (\ref{pendulum}) read:
\begin{align}
\mathcal{F}_{\epsilon^3} &= (\Fethree+4\Fithree-2\Fei)/9, \\
\Upsilon_0^{\epsilon^3} &= 3(2(F_{\rm{GR}} + \Feone + g_5) - 4(\Fione + f_2) + (2 \Fethree  \nonumber \\
&- 4 \Fei - 4 \Fithree ) \bar{\mathcal{V}} ))(4\Fethree -8 \Fei + 16 \Fithree)^{-1} \nonumber ,
\end{align}
where $\bar{\mathcal{V}}$ is again evaluated at nominal actions.

The separatrix of the $(\phi-2\psi)$ resonance is shown as a cyan curve on Figure (\ref{Psection}). As can be gathered from the Figure, the homoclinic curves of $(\phi-\psi)$ and $(\phi-2\psi)$ resonances overlap in a nearly perfect fashion, insinuating chaotic motion throughout much of the domain covered by the critical curves \citep{Chirikov1979}. The rough agreement of the expected size and character of the chaotic layer with the numerical surface of section depicted in the same Figure suggests that these two resonances are indeed the ones primarily responsible for driving Mercury's stochastic motion. 

In order to check that no additional resonances of significant importance contribute to Mercury's chaotic evolution, we may take advantage of the flexible nature of our perturbative model and explore a regime where the resonances are not overlapped and the stochasticity parameter (defined below) is slightly below unity. To do this, we recompute the \Poincare\ surface of section shown in the left panel of Figure (\ref{Psection}), reducing the coupling constant\footnote{Recall that setting $\Fei = 0$ yields an integrable system.} $\Fei$ by a factor of $0.6$. The result is shown in the right panel of Figure (\ref{Psection}).

In this surface of section, most of the plotted orbits are quasi-periodic and the underlying resonant structure is clearly visible. The neighborhoods of the unperturbed separatricies of the $(\phi-\psi)$ and $(\phi-2\psi)$ resonances remain chaotic and are depicted in the Figure with black points. Additionally, there exist several chains of high-order resonances. However, because of their small widths they are unlikely to contribute to large-scale chaotic evolution significantly. This suggests that accounting for the $(\phi-\psi)$ and $(\phi-2\psi)$ resonances alone is sufficient to describe Mercury's chaotic evolution to a satisfactory approximation.

\subsection{Characteristic Lyapunov Time and Diffusive Transport in Action Space}

Generally, the eigenfrequency associated with the equilibria of the Hamiltonian (\ref{pendulum}) is given by 
\begin{align}
\lambda_{\epsilon^n} = n\sqrt{2\mathcal{F}_{\epsilon^n} \mathcal{C}_{\epsilon^n}},
\label{lambda}
\end{align}
while the resonance half-width reads:
\begin{align}
\Delta \Upsilon_{\epsilon^n} = \sqrt{\frac{2 \mathcal{C}_{\epsilon^n}}{\mathcal{F}_{\epsilon^n}}}.
\label{halfwidth}
\end{align}

With the above constants specified in the previous sub-section, the distance between the resonant equilibria is now also well defined:
\begin{align}
\delta \Upsilon = |\Upsilon_0^{\epsilon^2} - \Upsilon_0^{\epsilon^3}|.
\end{align}
The stochasticity parameter \citep{Yellowbook,MurrayHolman97} is concurrently defined as the ratio of the average resonance width to the distance between resonances:
\begin{align}
\mathcal{K} \equiv \frac{\langle \Delta \Upsilon \rangle}{\delta \Upsilon} = \frac{\Delta \Upsilon_{\epsilon^2} + \Delta \Upsilon_{\epsilon^3}}{2|\Upsilon_0^{\epsilon^2} - \Upsilon_0^{\epsilon^3}|}.
\end{align}
For the parameters relevant to Mercury, we obtain a stochasticity parameter of order unity: $\mathcal{K} \sim 1.05$, signaling marginal resonance overlap\footnote{This is consistent with the picture outlined in the surface of section (\ref{Psection}).}.

A mathematically equivalent way to treat the overlap of multiple nonlinear resonances is to view the chaotic layer as being periodically swept by a single separatrix  \citep{1985PhR...121..165E,1986PhRvA..34.4256C,1991PhyD...54..135H}. In a regime characterized by a stochasticity parameter close to unity, the separatrix sweeping period is of order the characteristic libration period, $2\pi/\lambda$ \citep{2001Natur.410..773M,Morbybook}. 

The time interval between successive encounters with the separatix is intimately related to the characteristic decoherence time of a bundle of nearby trajectories, or the Lyapunov time \citep{HolmanMurray1996,MurrayHolman97}. Accordingly, taking advantage of the fact that $\mathcal{K}\sim 1$ for the system at hand, we approximate the Lyapunov time as the inverse of the average of the unstable eigenvalues of the $(\phi-\psi)$ and $(\phi-2\psi)$ resonances:
\begin{align}
\tau_{\rm{L}} \sim \frac{1}{2}\frac{2\pi}{\langle \lambda \rangle}
\label{Lyap}
\end{align}
where the factor of $1/2$ accounts for the fact that the chaotic layer gets swept by the separatrix twice per libration period. The functional form of this expression is consistent with the estimate obtained by \citet{HolmanMurray1996} for the Asteroid belt. Quantitatively, equation (\ref{Lyap}) evaluates to $\tau_{\rm{L}} \sim 1.4$ Myr, in good agreement with $\tau_{\rm{L}}$ obtained from numerical integration of the perturbative model (\ref{HtwoDOF}) and published simulations \citep{Laskar89,SussmanWisdom1992,BatLaugh}.

On timescales significantly longer that the Lyapunov time, it is not sensible to imagine the evolution of a single orbit as representative. Instead, it is more sensible to consider the statistical properties of the evolution of the actions. Within a perfectly chaotic layer, the transport in action space is governed by the Fokker-Plank equation \citep{1945RvMP...17..323W}. For Hamiltonian systems, it can be shown that the Fokker-Plank equation simplifies to the diffusion equation \citep{1937PhRv...52.1251L} and the evolutionary properties of the system are captured by the chaotic diffusion coefficient, $\mathcal{D}$.

An upper bound on the diffusive excursion in action can be obtained by assuming the typical change in action to be of order the average half-width of the overlapped resonances while the characteristic timescale for such an excursion is the decoherence (or Lyapunov) time:
\begin{align}
\mathcal{D}_{\Upsilon} \lesssim \frac{\langle \Delta \Upsilon \rangle^2}{\tau_{\rm{L}}}.
\label{Dmax}
\end{align}

A somewhat better approximation for the quasi-linear diffusion coefficient may be obtained directly from the equations of motion. Specifically, following \citet{MurrayHolman97}, we employ the random-phase approximation\footnote{We additionally take the fraction of time spent in either resonance to be comparable.} \citep{Yellowbook} and estimate:
\begin{align}
\mathcal{D}_{\Upsilon} &\simeq \frac{1}{2\pi} \bigg\langle \int_0^{2\pi} \frac{\left( \tau_{\rm{L}} \mathcal{C}_{\epsilon^n} \sin(n \upsilon) \right)^2}{\tau_{\rm{L}}} d\upsilon \bigg\rangle \nonumber \\ 
&= \frac{\pi^2}{4 \tau_{\rm{L}}} \bigg\langle \frac{\mathcal{C}_{\epsilon^n}}{n^2 \mathcal{F}_{\epsilon^n}} \bigg\rangle  \sim \frac{\pi^2}{48}\frac{\langle \Delta \Upsilon \rangle^2}{\tau_{\rm{L}}}
\label{Dave}
\end{align}
With nominal parameters, we obtain $\mathcal{D}_{\Upsilon} \simeq 1.7 \times 10^{-10} (\Lambda/2)^2$ as an average estimate and $\mathcal{D}_{\Upsilon}^{\rm{max}} \simeq 8.4 \times 10^{-10} (\Lambda/2)^2$ as an upper bound on the diffusion coefficient.

Evaluation of the diffusion coefficient by numerical integration of the equations of motion (see Figure \ref{Psection}) yields $\mathcal{D}_{\Upsilon} = 1.74 \times 10^{-10} (\Lambda/2)^2$ and $\mathcal{D}_{\Upsilon}^{\rm{max}} = 7.63 \times 10^{-10} (\Lambda/2)^2$ for the average and maximum values respectively. These estimates are in excellent agreement with those obtained from equations (\ref{Lyap}), (\ref{Dmax}) and (\ref{Dave}). Consequently, we conclude that despite the approximations made in deriving the analytical results, the obtained values remain quantitatively sound. 

\section{The Onset of Large-Scale Instability}

\begin{figure}[t]
\centering
\includegraphics[width=1\columnwidth]{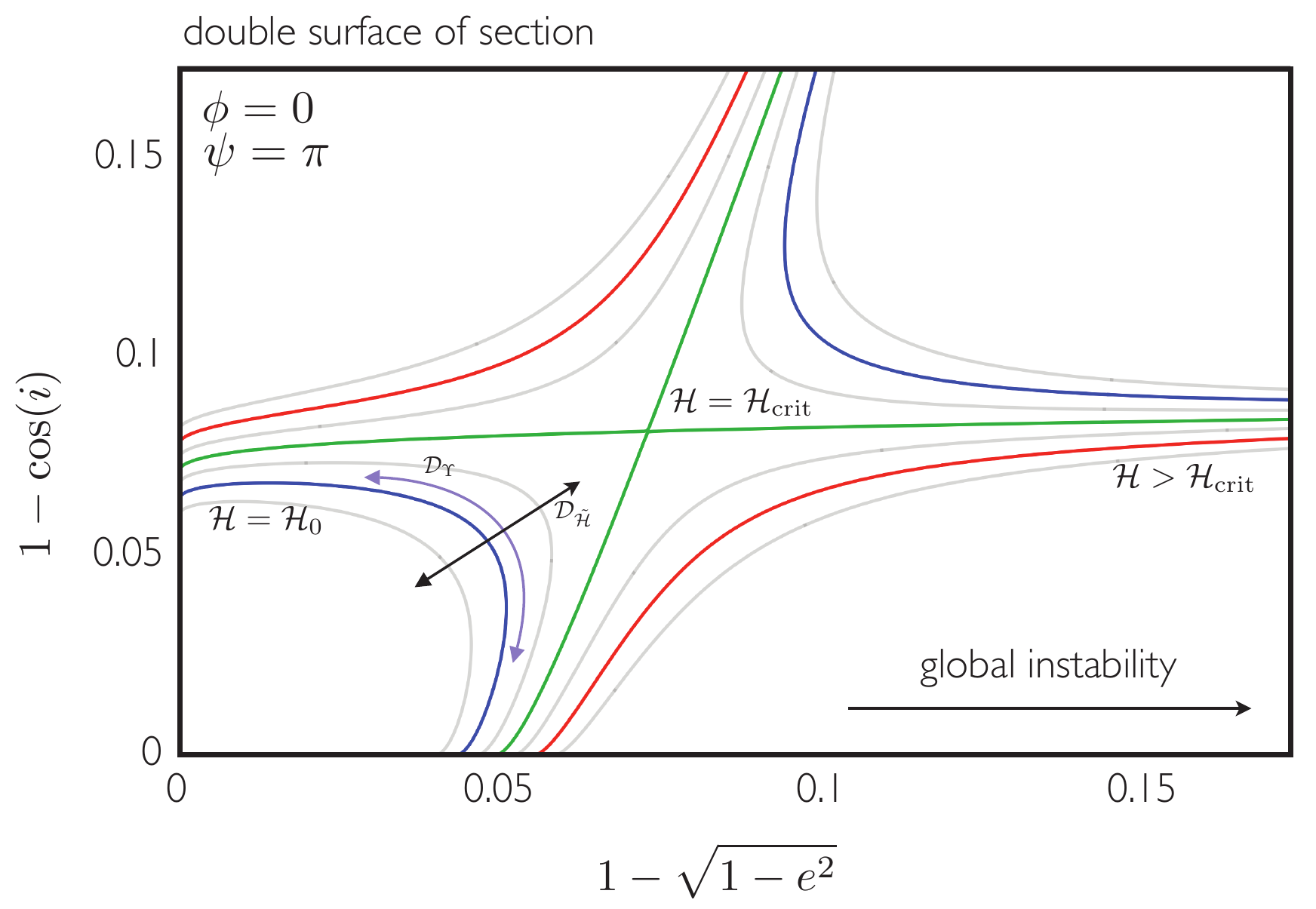}
\caption{A double section of $\mathcal{H}$ at $\phi=0,\psi=\pi$. The loci denote level curves of the Hamiltonian. The blue curve corresponds to the nominal value of $\mathcal{H}$ and is labeled $\mathcal{H}_0$. Note that there are two solutions on the double section for $\mathcal{H}=\mathcal{H}_0$: at present, Mercury resides on the stable solution, shown on the left-bottom quadrant of the figure. The critical locus (labeled $\mathcal{H}=\mathcal{H}_{\rm{crit}}$) is shown in green. A transition to the critical locus breaks the topological boundary between highly eccentric orbits and the present dynamical state, allowing the system to evolve towards a globally unstable configuration. As discussed in the text, it can also be shown that a transition through the critical locus corresponds to a transition through the $\nu_5$ ($\phi$) and $\nu_{12}$ ($\psi$) secular resonances. An $\mathcal{H}$-level significantly exceeding $\mathcal{H}_{\rm{crit}}$ is shown in red. Chaotic (Chirikov) diffusion along a locus arises from the overlap of high-order secular resonances (see Figure \ref{Psection}) and is labeled $\mathcal{D}_{\Upsilon}$. Diffusion across $\mathcal{H}$ levels is stochastically pumped by the irregular nature of the eccentricity evolution, and is labeled $\mathcal{D}_{\mathcal{H}}$ in the Figure. Generally, the transition to instability (i.e. diffusion of $\mathcal{H}$) occurs on a timescale much longer than that corresponding to the ergodic exploration of a given locus (i.e. diffusion of $\Upsilon$).}
\label{Dsection}
\end{figure}

The above-defined diffusion coefficient $\mathcal{D}_{\Upsilon}$ characterizes stochastic dispersal in Mercury's eccentricity and inclination within the admissible region of phase space. However, we have already argued that orbit crossing is not possible within the framework of the Hamiltonian (\ref{HtwoDOF}) because the admissible domain of the dynamics does not extend to a sufficiently high eccentricity. What additional ingredient is needed for the simplified system to successfully exhibit bounded-unbounded chaotic transitions? An answer to this question can be gathered by constructing a double surface section of the Hamiltonian (\ref{HtwoDOF}). Figure (\ref{Dsection}) depicts such a double section, where the angles of the Hamiltonian have been fixed at $\phi=0$ and $\psi=\pi$ and multiple levels of $\mathcal{H}$ are plotted.

As stochastic evolution carries the trajectory through the chaotic sea, every time the angles $\phi$ and $\psi$ randomly line up to $\phi=0$ and $\psi=\pi$, the actions of the orbit will map onto the blue locus labeled $\mathcal{H} = \mathcal{H}_0$ on Figure (\ref{Dsection}). In other words, as chaotic evolution proceeds, the system ergodically explores the portion of the blue locus not occupied by quasi-periodic trajectories\footnote{We note that for the same value of $\mathcal{H} = \mathcal{H}_0$, there exists another admissible locus in Figure (\ref{Dsection}) that resides at high eccentricity and inclination. This solution is characterized by prograde rather than retrograde circulation of $\nu$ and transitions between the loci is not permitted as long as $\mathcal{H}$ remains conserved.}. 

Let us now imagine that the value of $\mathcal{H}$ is slowly modulated towards a critical value $\mathcal{H} \rightarrow \mathcal{H}_{\rm{crit}}=1.17\ \mathcal{H}_0$, shown as a green curve on Figure (\ref{Dsection}). As long as $\mathcal{H} < \mathcal{H}_{\rm{crit}}$, chaotic diffusion of Mercury's orbit remains bounded because the loci delineated on Figure (\ref{Dsection}) are qualitatively similar to the nominal $\mathcal{H} = \mathcal{H}_0$ curve. However, a drastically different turn of events can be envisioned if the value of $\mathcal{H}$ is allowed reach $\mathcal{H}_{\rm{crit}}$. Indeed, at $\mathcal{H} = \mathcal{H}_{\rm{crit}}$, the topology of the double section changes, such that the loci connect to catastrophic values of the actions. Consequently, if the dynamics remains globally chaotic at $\mathcal{H} \geqslant \mathcal{H}_{\rm{crit}}$, the system will diffusively evolve towards orbit crossing on a timescale considerably grater than, but nevertheless comparable to the Lyapunov time. 

The value of the Hamiltonian $\mathcal{H}=\mathcal{H}_{\rm{crit}}$ holds additional physical meaning beyond being a simple topologic transition in phase-space. Particularly, it is a locus that corresponds to null oscillation frequencies\footnote{This naturally follows from Hamilton's equations.} of the angles $\phi$ and $\psi$. In other words, the corresponding green curve depicted on the double section (\ref{Dsection}) tracks the locations of the $\nu_5$ and $\nu_{12}$ secular resonances (i.e. resonances associated with the leading order critical angles $\phi$ and $\psi$). Consequently, the evolution of $\mathcal{H}$ towards $\mathcal{H_{\rm{crit}}}$ within the framework of our model is equivalent to the evolution of the system towards linear secular resonance, as observed in numerical experiments \citep{Laskar08,BatLaugh}.

This discussion highlights the additional component needed to fully capture Mercury's orbital evolution on multi-Gyr timescales: the conservation of $\mathcal{H}$ must be broken. Consequently, it is natural to infer that the process that governs the transition between bounded and unbounded chaos in Mercury's case is the slow diffusion of $\mathcal{H}$ itself.
Moreover, an examination of the double section (\ref{Dsection}) intuitively explains why the onset of large-scale instability appears to occur ``suddenly" in numerical simulations.

\subsection{A Chaotic Model With 2.5 Degrees of Freedom}

The Hamiltonian itself, $-\mathcal{H}$, is an action conjugate to $t$. Thus, in order to capture the transitions between bounded and unbounded chaos, we must incorporate explicit time-dependence into the governing equations. To do this, we retain an additional term in the decomposition (\ref{Venussolution}), which generates an extra harmonic in the Hamiltonian. Specifically, we shall retain the $g_2$ mode, as it is the largest amplitude, slowly varying term that remains in the expansion\footnote{We note that although \citet{LithWu11} did not elucidate the role of the harmonic associated with this mode in their investigation, they did point out that its inclusion into the Hamiltonian had an unknown but important effect on the dynamics.}.

Applying the transformations (\ref{transformone}) yields the following expression:
\begin{empheq}[box=\fbox]{align}
\tilde{\mathcal{H}} &= \left(F_{\rm{GR}} + \Feone + g_5 \right) \Phi + \Fethree \Phi^2  \nonumber \\
&+ \Fetwofive \sqrt{\Phi} \cos(\phi) + \left(\Fione + f_2 \right) \Psi + \Fithree \Psi^2 \nonumber \\ 
&+ \Fitwo \sqrt{\Psi} \cos(\psi) + \Fei \Phi \Psi \nonumber \\
&+ \Fetwotwo \sqrt{\Phi} \cos(\phi + (g_2 - g_5)t + (\beta_2 - \beta_5)).
\label{HthreeDOF}
\end{empheq}
In contrast with equation (\ref{HLLphipsi}), in Hamiltonian (\ref{HthreeDOF}) we have chosen not to extend the phase space and thus retain the non-autonomous nature (denoted by a tilde) of $\tilde{\mathcal{H}}$ \citep{Morbybook}. The newly introduced constant, in some similarity with equations (\ref{coeffsF}), reads:
\begin{align}
\Fetwotwo &= -\sum_{j = 2}^8 \left( \frac{\mathcal{G} m m_j}{a_j} \sqrt{\frac{2}{\Lambda}} \fetwoj \bar{e}_{j,2} \right) \nonumber \\
&= -1.72 \times 10^{-10}.
\label{coeffsF4}
\end{align}

\begin{figure}[t]
\centering
\includegraphics[width=1\columnwidth]{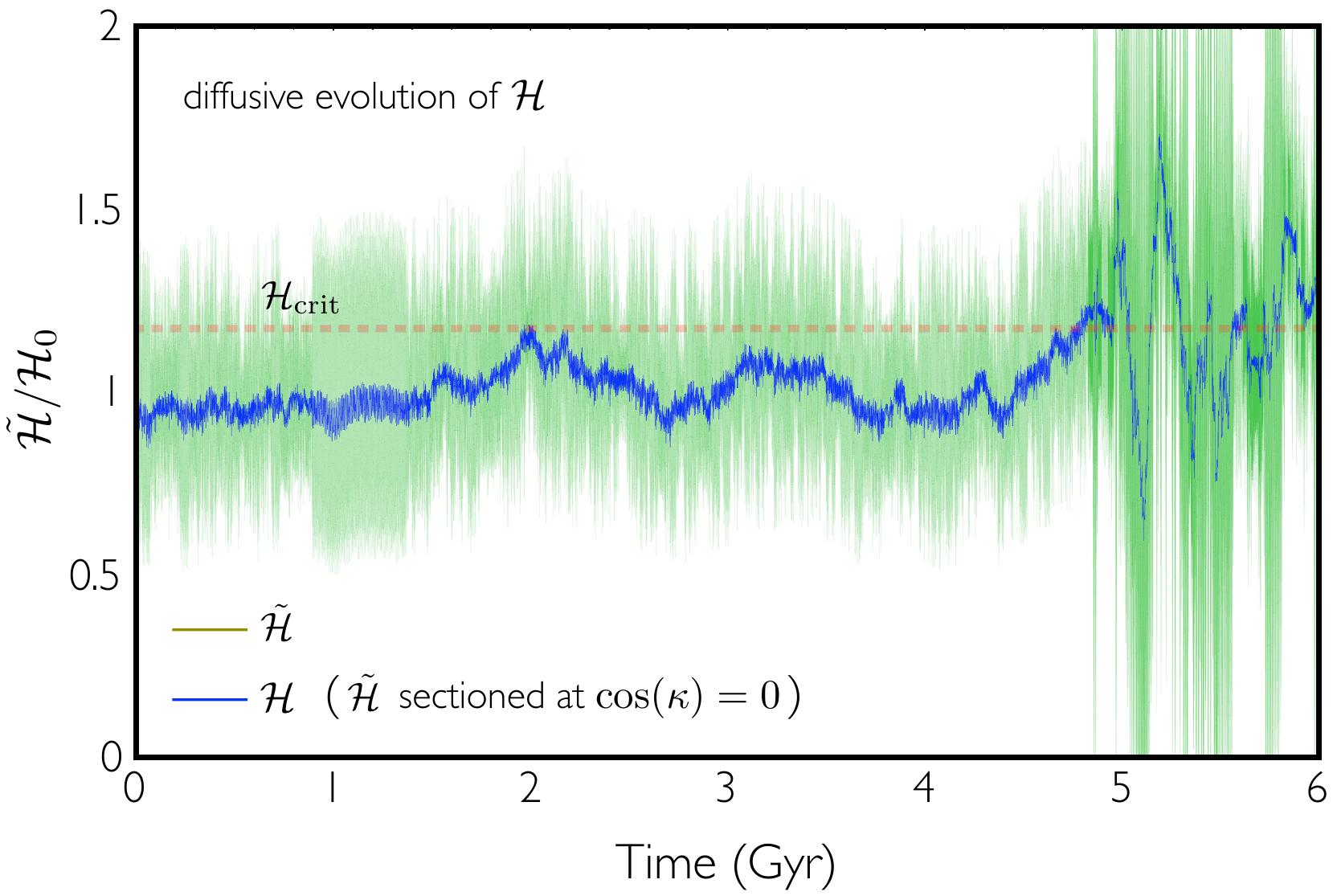}
\caption{Stochastic evolution of $\tilde{\mathcal{H}}$. The presented evolutionary sequence was obtained within the framework of the non-autonomous Hamiltonian (\ref{HthreeDOF}) and corresponds to the orbital solution showed as a blue curve in Figure (\ref{ei}). The green curve corresponds to a direct evaluation of equation (\ref{HthreeDOF}) at an arbitrary cadence. Concurrently, the blue curve denotes $\tilde{\mathcal{H}}$, evaluated whenever the time-dependent harmonic $\kappa  = \phi + (g_2 - g_5)t + (\beta_2 - \beta_5)$ crosses its initial value of $\kappa = \pi/2$. Accordingly, the blue curve tracks the average value of the non-autonomous system, yielding a closer correspondence to its autonomous counterpart. Note that large-scale instability is triggered as a consequence of the intersection of the average value of $\tilde{\mathcal{H}}$ with $\mathcal{H}_{\rm{crit}}$, as suggested by Figure (\ref{Dsection}).}
\label{Hoft}
\end{figure}

With explicit time-dependence in place, the slow diffusion of $\tilde{\mathcal{H}}$ can indeed carry the trajectory to catastrophically high eccentricity. This is elucidated in Figure (\ref{ei}) where the evolution stemming from the same nominal initial conditions invoked before, but governed by the Hamiltonian (\ref{HthreeDOF}), is shown in blue. In stark contrast with the trajectory obtained from the autonomous model (\ref{HtwoDOF}), in this numerical experiment Mercury successfully evolves into the $\nu_5$ and $\nu_{12}$ secular resonances and becomes violently unstable on a timescale comparable to the remaining main-sequence lifetime of the Sun. 

With a numerical solution at hand, it is possible to check the sensibility of the qualitative discussion regarding the approach of $\mathcal{H}$ towards $\mathcal{H}_{\rm{crit}}$ as the cause of the instability, quoted above. To do this, let us examine the evolution of the value of $\tilde{\mathcal{H}}$ as a function of time, shown in Figure (\ref{Hoft}). The green curve depicted in the Figure represents the value of $\tilde{\mathcal{H}}$ (normalized by its initial value) as given by equation (\ref{HthreeDOF}) and sampled at an arbitrary cadence in time. As can be seen, this function crosses $\mathcal{H}_{\rm{crit}}$ repeatedly before the onset of instability at $\sim 5$Gyr. However, it should be understood that there is not a one-to-one correspondence between $\tilde{\mathcal{H}}$ and $\mathcal{H}$ since the former incorporates an additional, time-dependent harmonic. This harmonic yields rapid oscillations in the value of the Hamiltonian, obscuring a candid comparison between the evolution of $\tilde{\mathcal{H}}$ and the double section of $\mathcal{H}$. 

A more sensible comparison can be made by sectioning the evolution on the time-dependent angle $\kappa = \phi + (g_2 - g_5)t + (\beta_2 - \beta_5)$, and plotting $\tilde{\mathcal{H}}$ only when $\kappa$ corresponds to its initial value, which we set to $\pi/2$, such that the initial values of $\tilde{\mathcal{H}}$ and $\mathcal{H}$ are also identical. This procedure effectively tracks the average value of $\tilde{\mathcal{H}}$ and is shown as a blue curve on Figure (\ref{Hoft}). With a more direct connection between $\tilde{\mathcal{H}}$ and $\mathcal{H}$ established, it is immediately apparent that the onset of instability corresponds to a point when the average value of $\tilde{\mathcal{H}}$ crosses $\mathcal{H}_{\rm{crit}}$, as discussed above.

With all of the desired effects (bounded chaotic diffusion on a $\sim$ Myr timescale and the onset of global instability on a $\sim$ Gyr timescale) accounted for, Hamiltonian (\ref{HthreeDOF}) constitutes the simplest dynamical model for Mercury's secular evolution. Because $\tilde{\mathcal{H}}$ is based on a classical series expansion of the disturbing gravitational potential (which treats $e$ and $s$ as small parameters; \citealt{MD99}), its quantitative agreement with a full N-body model (see e.g. \citealt{Quinn1991,SussmanWisdom1992}) should not be expected to be superb. However, as we show below, the characteristic dynamical lifetime derived from $\tilde{\mathcal{H}}$ is in moderately good agreement with numerical experiments \citep{Laskar08,BatLaugh,Laskar09} and reproduces the qualitative behavior of the solutions well.

\subsection{Chaotic Diffusion of $\mathcal{H}$ and the Dynamical Lifetime of the Solar System}

Although the Hamiltonian (\ref{HthreeDOF}) captures the onset of large-scale instability and the Hamiltonian (\ref{HtwoDOF}) does not, as long as the system resides in the bounded chaotic regime, the stochastic properties of the orbits governed by the two models (i.e. $\tau_{\mathcal{L}},\mathcal{D}_{\Upsilon}$) are nearly indistinguishable. Consequently, with the chaotic properties of the two degree of freedom model delineated in the previous section, we are now in a position to estimate the global lifetime of the system by considering the chaotic diffusion of $\mathcal{H}$. The relevant equation of motion, stemming from Hamiltonian (\ref{HthreeDOF}) reads:
\begin{align}
\frac{d \mathcal{H}}{dt} &= \{\mathcal{H},\tilde{\mathcal{H}}\} = \Fetwotwo \Fetwofive \sin(\omega t + \Delta\beta_{2,5}) \nonumber \\
&+ \Fetwotwo \sqrt{\Phi} \left((F_{\rm{GR}}+ \Feone +g_5) + 2 \Fethree \Phi + \Fei \Psi \right)\nonumber \\
&\times \sin(\phi+ \omega t + \Delta\beta_{2,5}).
\label{dHdtPB}
\end{align}

The leading term in the above expression does not induce any long-term drift in $\mathcal{H}$ and can therefore be ignored. Moreover, recall that the basis of the series expansion of the Hamiltonian is an assumption of small eccentricities and inclinations \citep{2000Icar..147..129E}. Thus, for tractability we may discard terms of superior order in the actions, as their effects will be secondary. An approximate expression for variation in $\mathcal{H}$ then reads:
\begin{align}
\frac{d \mathcal{H}}{dt} &\simeq  \Fetwotwo (F_{\rm{GR}}+ \Feone +g_5) \sqrt{\Phi} \nonumber \\
&\times \sin(\phi + \omega t + \Delta\beta_{2,5}),
\label{dHdt}
\end{align}
where $\omega = (g_2 - g_5)$ and $\Delta\beta_{2,5}$ is a phase constant which we set to $\pi/2$, as above (this does not change the results in any meaningful way).

From the form of equation (\ref{dHdt}), it is immediately clear that if the evolution of $\Phi$ and $\phi$ is forced to be strictly periodic, the evolution of $\mathcal{H}$ will be quasi-periodic. Consequently, in absence of low-order secular resonances between the angles $\phi,\psi$ and $\omega t$, it is natural to treat the chaotic evolution of $\mathcal{H}$ as if it is driven ``extrinsically" by the chaotic properties of the $(\Phi,\phi)$ degree of freedom, effectively reducing equation (\ref{dHdt}) to the Langevin stochastic differential equation \citep{Stochastic}. Such calculations are often referred to as chaotic pump calculations of Arnold diffusion \citep{1971JSP.....3..307C, Yellowbook}. 

An important caveat inherent to this procedure of breaking up the Hamiltonian into two parts is the assumption that the changes in $\tilde{\mathcal{H}}$ (prior to large-scale instability) do not affect the dynamics of the remaining degrees of freedom significantly. In the case of Mercury this assumption holds, however it is important to keep in mind that it need not generally. In the contrary case, one would proceed to calculate the diffusion of $\tilde{\mathcal{H}}$ in a conventional Chirikov fashion, as done above.

As a starting assumption of the calculation, we note that the characteristic frequency of the time-dependence in $\tilde{\mathcal{H}}$ greatly exceeds the Lypunov exponent:
\begin{align}
\tau_{\rm{L}} \gg 2\pi/\omega.
\end{align}
This means that over a single cycle of the critical angle in equation (\ref{dHdt}), the evolution of $\Phi$ will appear only mildly stochastic. Thus, over the relevant timescale, we can approximate the time-evolution of $\Phi$ as being composed of a periodic (Lagrange-Laplace) component and a smaller stochastic component:
\begin{align}
\sqrt{\Phi} = \sqrt{\Phi_{\rm{LL}}+\Phi_{\rm{st}}} \simeq \sqrt{\Phi_{\rm{LL}}} + \frac{\Phi_{\rm{st}}}{2 \sqrt{\langle \Phi_{\rm{LL}} \rangle}}.
\label{decompositionstuff}
\end{align}

Equation (\ref{dHdt}) is now composed of two terms\footnote{Note that the denominator of the second term in equation (\ref{decompositionstuff}) is set to the time-average of the periodic component of the solution for simplicity.}. The term multiplied by $\sqrt{\Phi_{\rm{LL}}}$ leads to rapid short-term variations of $\mathcal{H}$ and has no appreciable long-term contribution. Consequently, to capture the diffusive property of $\tilde{\mathcal{H}}$, we can concentrate entirely on the stochastic part:
\begin{align}
\bigg\langle\frac{d \mathcal{H}}{dt} \bigg\rangle &=  \Fetwotwo (F_{\rm{GR}}+ \Feone +g_5) \frac{\Phi_{\rm{st}}}{2 \sqrt{\langle \Phi_{\rm{LL}} \rangle}} \nonumber \\
 &\times \big(\cos(\phi)\cos(\omega t) - \sin(\phi)\sin(\omega t) \big),
\label{dHdt2}
\end{align}

Over timescales that are short compared to the Lyapunov time the random-phase approximation does not apply to $\phi$. This means that although the integral of equation (\ref{dHdt2}) will exhibit some short-term variation, it will be nearly periodic. However, on a timescale of order a few Lyapunov times, the cartesian components of the eccentricity vector $\sqrt{2 \Phi_{\rm{st}}} \cos(\phi)$ and $\sqrt{2 \Phi_{\rm{st}}} \sin(\phi)$ act as independent uncorrelated random variables. An alternative viewpoint is that it takes of order a Lyapunov time to build up the inherent randomness of the integral. Thus, an integral of equation (\ref{dHdt2}) over a characteristic de-correlation time will amount to an integral over a single cycle of the critical angle under the de-correlated assumption. 

Moreover, it is sensible to assume that averaged over timescales longer than $\tau_{\rm{L}}$,
\begin{align}
\langle \Phi_{\rm{st}} \cos(\phi)\cos(\omega t) \rangle \sim \langle \Phi_{\rm{st}} \sin(\phi)\sin(\omega t) \rangle
\end{align}
because the diffusive properties of the components of the eccentricity vector are the same. With the aforementioned arguments in mind, we may write
\begin{align}
\int_0^{\tau_{\rm{L}}} \bigg\langle\frac{d \tilde{\mathcal{H}}}{dt} \bigg\rangle dt &\sim \frac{\Fetwotwo (F_{\rm{GR}}+ \Feone +g_5)}{\sqrt{\langle \Phi_{\rm{LL}} \rangle}} \nonumber \\
&\times \int_0^{2\pi/\omega}\Phi_{\rm{st}} \sin(\phi) \sin(\omega t) dt.
\label{intH}
\end{align}
The above expression implies that the evolution of $\mathcal{H}$ can be envisioned as a random walk with a characteristic step size given by the quoted stochastic integral over a single circulation cycle of $\omega$ and a characteristic step time of order a Lyapunov time.

It now remains only to evaluate the stochastic integral (\ref{intH}). As a leading order approximation, let us assume that the evolution of $\Phi_{\rm{st}} \sin(\phi)$ is akin to drift-free Weiner process, $\mathcal{W}$, with the standard deviation set to the chaotic diffusion coefficient\footnote{If $e$ and $s$ behave like gaussian random variables with similar variance, then $\mathcal{D}_{\Phi} \sim \mathcal{D}_{\Upsilon}/\sqrt{2}$, however such factors of order unity are unimportant at the level of approximation employed in this work.}, $\mathcal{D}_{\Upsilon}$ \citep{probabilitybook}. 

Let $\varphi = \omega t$. We now have:
\begin{align}
\Delta \mathcal{H} &= \frac{1}{\omega} \sqrt{\frac{ \mathcal{D}_{\Upsilon}}{\omega} }\frac{\Fetwotwo (F_{\rm{GR}}+ \Feone +g_5)}{\sqrt{\langle \Phi_{\rm{LL}} \rangle}} \nonumber \\
&\times \int_0^{2\pi}\mathcal{W}_{\varphi}\sin(\varphi) d\varphi,
\label{DeltaH}
\end{align}
where $\mathcal{W}_{\varphi}$ is $\mathcal{W}$, scaled such that the time unit is dimensionless (i.e. $\omega t$).

Next, we note that 
\begin{align}
d (\mathcal{W}_{\varphi} \cos(\varphi)) = (d\mathcal{W}_{\varphi}) \cos(\varphi) - \mathcal{W}_{\varphi} \sin(\varphi) d\varphi.
\end{align}
In turn, this means that 
\begin{align}
&\int_0^{2\pi}\mathcal{W}_{\varphi}\sin(\varphi) d\varphi = \int_0^{2\pi} \cos(\varphi) d\mathcal{W}_{\varphi} - \mathcal{W}_{\varphi} \cos(2 \pi) \nonumber \\
&= \int_0^{2\pi} \left(\cos(\varphi) - \cos(2\pi) \right) d\mathcal{W}_{\varphi}.
\label{molei}
\end{align}

The integral (\ref{molei}) is thus a Gaussian random variable with zero mean and variance \citep{Stochastic}:
\begin{align}
\int_0^{2\pi} \left(\cos(\varphi) - \cos(2\pi) \right)^2 d \varphi = 3 \pi.
\end{align}
Putting this result together with expression (\ref{DeltaH}), we obtain the following estimate for the diffusion coefficient of $\tilde{\mathcal{H}}$:
\begin{align}
\mathcal{D}_{\mathcal{H}} = \mathcal{D}_{\Upsilon} \left(\frac{ 3 \pi}{\omega^3 \tau_{\rm{L}}} \frac{(\Fetwotwo (F_{\rm{GR}}+ \Feone +g_5))^2}{\langle \Phi_{\rm{LL}} \rangle} \right).
\label{DH}
\end{align}

Accordingly, given a value $\mathcal{H}_{\rm{crit}}$, at which the system transitions to global instability and an initial condition $\mathcal{H}_0$, we may estimate the characteristic dynamical lifetime of the Solar System as:
\begin{align}
\mathcal{T} \sim \frac{\left(\mathcal{H}_{\rm{crit}}-  \mathcal{H}_0 \right)^2}{\mathcal{D}_{\mathcal{H}}}.
\label{lifetime}
\end{align}
Quantitatively, expression (\ref{lifetime}) evaluates to $\mathcal{T} \sim 10^8 \-- 10^9$ years. This estimate is in good agreement with repeated numerical integration of the Hamiltonian (\ref{HthreeDOF}) with slightly different initial conditions but is considerably shorter than the typical dynamical lifetimes obtained by precise numerical models (e.g. \citealt{Laskar09}). 

\begin{figure}[t]
\centering
\includegraphics[width=1\columnwidth]{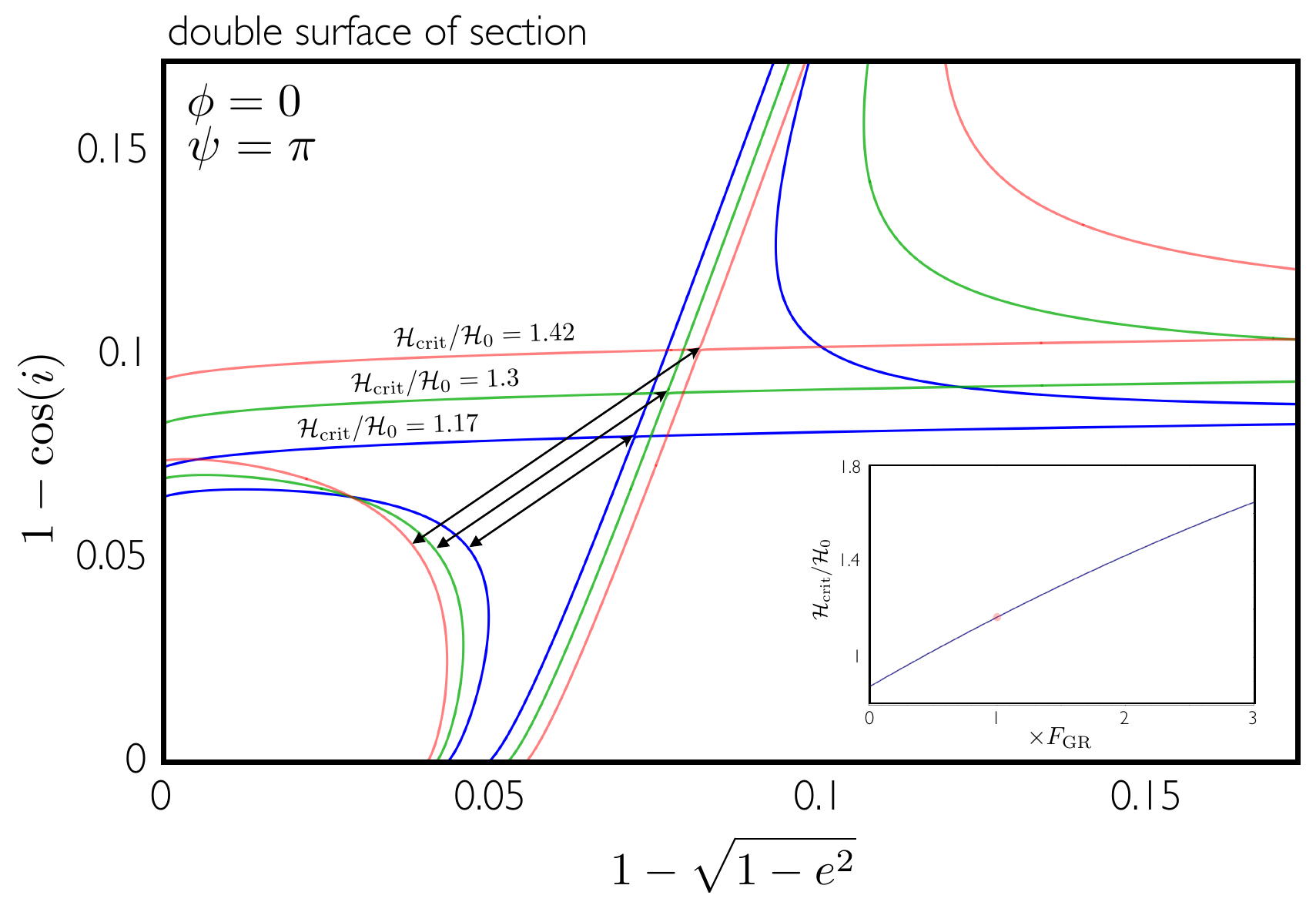}
\caption{Effects of general relativity on the stability of the Solar System. In some parallel with Figure (\ref{Dsection}), this Figure shows loci corresponding to nominal and critical values of $\mathcal{H}$ for various degrees of amplification of the relativistic precession. As the relativistic contribution to the apsidal precession of Mercury is increased, so is the difference between the nominal and the critical values of the Hamiltonian. Specifically, the normalized value of $\mathcal{H}_{\rm{crit}}$ increases from its nominal value of $\mathcal{H}_{\rm{crit}}/\mathcal{H}_0 = 1.17$ to $\mathcal{H}_{\rm{crit}}/\mathcal{H}_0 = 1.3$ and $\mathcal{H}_{\rm{crit}}/\mathcal{H}_0 = 1.42$, as the relativistic correction is enhanced by factors of $1.5$ and $2$ respectively (further amplification significantly alters $\mathcal{D}_{\Upsilon}$ obscuring candid interpretation). Moreover as shown in the inset, if the relativistic correction is neglected within the context of our simplified model, $\mathcal{H}_{\rm{crit}}/\mathcal{H}_0$ drops below unity meaning that the system becomes unstable immediately upon initiation.}
\label{GR}
\end{figure}

Reasons for this almost certainly arise from the imperfect nature of our simplified secular model. Firstly, as already pointed out above, the characteristic Lyapunov time obtained within the context of our treatment is somewhat shorter than that of the real Mercury \citep{Laskar89,SussmanWisdom1992}. Second, our model is based on a series expansion that treats $e$ and $i$ as small parameters, and therefore becomes increasingly imprecise as the critical value of $\mathcal{H}$ is approached. To this end, we have also adopted the Lagrange-Laplace solution as a description of the dynamical evolution of planets other than Mercury, which introduces additional inaccuracies. Finally, it is well known that the dynamical stability of the Solar System is sensitive to small changes in the underlying parameters and it is therefore not too surprising that somewhat different quantitive results are obtained provided distinct models. Nevertheless, the obtained instability timescale exceeds the Lyapunov time by more than two orders of magnitude and thus captures the long-term chaotic behavior of the inner Solar System well, on a qualitative level.

\subsection{General Relativistic Effects}

Among the more surprising features of the Solar System's dynamical behavior is its significant dependence on relativistic effects. The stabilizing role of general relativity was first noted in the works of \citet{Laskar08,BatLaugh} and explored more thoroughly in the study of \citet{Laskar09}. Specifically, the latter investigation determined that while a purely Newtonian Solar System has a $\sim 60\%$ probability of becoming unstable within the next $5$ Gyr, accounting for relativistically-induced apsidal precession of Mercury reduces the chances to $\sim 1\%$.

Although the quantitative agreement between the dynamical lifetime predicted by our simplified model and the Solar System's true dynamical lifetime is imperfect, it is still interesting to explore its dependence on underlying parameters. To do this, we manually enhance or diminish $F_{\rm{GR}}$ in the Hamiltonian and monitor the difference between the initial and catastrophic values of $\mathcal{H}$, i.e. $\left(\mathcal{H}_{\rm{crit}}-\mathcal{H}_0 \right)$. 

Figure (\ref{GR}) mirrors the double section depicted in Figure (\ref{Dsection}) and shows pairs of loci corresponding to $\mathcal{H}_0$ and $\mathcal{H}_{\rm{crit}}$ for nominal $F_{\rm{GR}}$ (blue), $F_{\rm{GR}}$ enhanced by a factor of 1.5 (green) and $F_{\rm{GR}}$ enhanced by a factor of 2 (pink). As can be gathered from the Figure, the separation in $\mathcal{H}$ between bounded and unbounded chaotic states increases quasi-linearly as the relativistically facilitated apsidal precession is enhanced\footnote{Another way to view this effect is that increasing $F_{\rm{GR}}$ de-tunes the system away from the $\nu_5$ and $\nu_{12}$ secular resonances (\citealt{BatLaugh}; see also \citealt{2006IJMPD..15.2133A}).}. Accordingly, expression (\ref{lifetime}) suggests that dynamical lifetime increases approximately as $\mathcal{T} \appropto (F_{\rm{GR}})^2$. 

This is in agreement with repeated numerical integration of the system governed by Hamiltonian (\ref{HthreeDOF}). Although, simulations also show that if $F_{\rm{GR}}$ is enhanced by a factor of $\sim3$ or greater, the dynamical lifetime shortens significantly. This behavior is likely associated with the changes in the diffusive properties of the two degree of freedom system (\ref{HtwoDOF}), which also enter into the expression for $\mathcal{T}$ through the diffusion coefficient $\mathcal{D}_{\mathcal{H}}$ (see equation \ref{DH}). Moreover, the possible appearance of high-order secular resonances between the three angles $\phi,\psi$ and $\omega t$ may dramatically accelerate the diffusion rate of $\tilde{\mathcal{H}}$ \citep{Yellowbook}.

An additional peculiar feature of our model is that if the relativistic correction is neglected entirely, at nominal parameters $\mathcal{H}_0$ exceeds $\mathcal{H}_{\rm{crit}}$, signaling immediate instability. This further highlights the fact that our perturbative model is by default closer to linear secular resonance than the real Solar System, in concurrence with a somewhat shorter derived dynamical lifetime quoted above.

Despite considerable limitations arising from a perturbative treatment of the gravitational interactions, the qualitative features of the Solar System's dynamical dependence on relativistic effects seems to be well-represented by the simple model at hand. Therefore, cumulatively it would appear that introducing additional complications into the calculations presented in this work is unlikely to yield further insights of considerable value. In other words, the analytical estimates derived from the perturbative model considered here probably capture the dominant characteristics of the Solar System's dynamical evolution in an acceptable manner.

\section{Discussion}

In this study, we have revisited the centuries-old question of the long-term dynamical evolution of the Solar System \citep{LaskarPoincare}, from an analytical perspective. We began by construing a simple Hamiltonian model based on a classical expansion of the gravitational potential \citep{LeVerrier1855,2000Icar..147..129E}, that successfully captures the stochastic, yet bounded character of Mercury's orbit on multi-Myr timescales. Building on the work of \citet{Laskar89,LithWu11,Boue} we applied canonical perturbation theory utilizing Lie transform methods \citep{1969CeMec...1...12D,1993CeMDA..55..101M} in order to elucidate the two primary high-order secular resonances that drive chaotic diffusion. While the overlap of non-linear secular resonances had already been conjectured to drive Mercury's stochastic motion (see e.g. \citealt{Laskar89,Laskar96,LithWu11}), this marks the first explicit identification of the specific angles primarily responsible for irregular dynamics.  

The overlap of the aforementioned resonances is in essence perfectly non-adiabatic \citep{Chirikov1959,Chirikov1979}. Accordingly, taking advantage of a near-unity stochasticity parameter (i.e. marginal overlap of the resonances), we utilized the perturbative model to analytically obtain the Lyapunov time and the chaotic diffusion coefficient \citep{Yellowbook} inherent to Mercury's orbit (see also \citealt{HolmanMurray1996,MurrayHolman97}). The resulting estimates are in good agreement with numerical determinations and qualitatively illustrate the origin and properties of Mercury's secular evolution. Moreover, the calculations are of considerable pedagogical value, as they demonstrate a successful reduction of a rather complex gravitational N-body problem to a tangible one. 

In a subsequent effort, we extended the model to account for transitions between bounded (stable) and unbounded (unstable) chaotic states, thereby illuminating the dynamical architecture that underlies the onset of large-scale instabilities. Specifically, we showed that the acquisition of catastrophic orbital parameters is facilitated by a topological transition in the structure of the governing Hamiltonian and it is the diffusion of the Hamiltonian itself that dictates the commencement of orbital disintegration. Accordingly, in this work we have provided the first purely analytical estimate of the inner Solar System's dynamical lifetime.

The characteristics of the slow diffusion inherent to Mercury's orbit bear some resemblance to Arnold diffusion \citep{Arnold1964,1978mmcm.book.....A,1971JSP.....3..307C,Yellowbook}. That is, initially the system resides in a particular high-order $(\phi-\psi)$ secular resonance that is perturbed by a yet higher order resonance, namely $(\phi-2\psi)$. In due course, as the value of $\tilde{\mathcal{H}}$ evolves stochastically the high-order resonant structure is maintained until the trajectory enters a distinct, doubly-resonant domain characterized by the overlap of the (lower-order) $(\phi)$ and $(\psi)$ resonances (see \citealt{1977RuMaS..32....1N}). Still, the parallel between the typically quoted Nekhoroshev structure (e.g. \citealt{1995JSP....78.1607M}) and the perturbative system described here is imperfect, as the former considers a single resonance perturbed by a higher-order remainder that is exponentially small in the perturbation parameter, where as the system at hand is ubiquitously chaotic. 

Of course, Mercury is not the only Solar System object whose orbit is susceptible to dynamical instabilities. For example, by now it is well know that there exist numerous unstable mean-motion resonant orbits within the Asteroid belt \citep{Wisdom80,Wisdom83, Morby90a, Morby90b}. Generally, the unstable resonances are devoid of objects, as chaotic diffusion of eccentricity allows asteroids to eventually acquire planet-crossing orbits and escape the Solar System (\citealt{1992AJ....104.1230L}). To this end, it is worthwhile to notice that the dynamical processes by which ejection is brought about for Solar System small bodies and planets are distinct. That is, asteroids are removed directly as a result of eccentricity diffusion (\citealt{MurrayHolman97}, see also \citealt{MurrayHolman99}), while Mercury's instability is triggered by a change in the dynamical structure of the Hamiltonian and the associated diffusion of the Hamiltonian itself. This difference highlights a certain diversity inherent to chaotic evolution and the emergence of instabilities in planetary systems.

Although the focus of this work weighs heavily on the dynamics of the inner Solar System, the developed model builds on the general Lagrange-Laplace secular theory \citep{MD99} and should be applicable to a wide array of planetary systems dominated by secular interactions \citep{2011ApJ...735..109W}. Indeed, Laplace-Lagrange theory has already found wide-spread applications to the study of the dynamics of extrasolar planetary systems \citep{2004Icar..168..237M,2007ApJ...661.1311V, 2009MNRAS.392....2M, 2009MNRAS.395.1777M} and the extension of the theory delineated here may play a significant role in clarifying the origins of the known extrasolar orbital architectures as well as their future evolutionary sequences.

The keen importance of understanding orbital instabilities in a general context is highlighted by the orbital distribution of extrasolar planets that do not reside in close proximity to their host stars. In particular, the continuous radial velocity and transit monitoring of the local galactic neighborhood has shown that severely excited orbits are not uncommon in typical planetary systems. In turn, this observational fact has been invoked as evidence for planet-planet scattering as a dominant mechanism responsible for sculpting the dynamical architectures \citep{JuricTremaine, FordRasio08, Chaterjee08, Raymond09a}. At present, the process by which newly formed planetary systems acquire unstable orbits remains elusive \citep{2011CeMDA.111..219B, 2013MNRAS.431.3494L} and an analysis much like the one performed in this work may be required to shed light on the early dynamical transmutation of orbital states. Consequently, the theoretical analysis performed herein is likely to find broad-ranging consequences for the interpretation of instability-driven dynamical evolution of generic planetary systems.

\acknowledgments
\textbf{Acknowledgments}. We are grateful to Norm Murray, Greg Laughlin, Fred Adams, Gongjie Li, and Daniel Tamayo for useful discussions. Additionally, we are thankful to Molei Tao for sharing his expertise in stochastic calculus with the authors. Finally, we wish to thank the anonymous referee, whose insightful report led to a substantial improvement of the paper.  

\begin{appendix}

\section{Constants of the Hamiltonian}
The constants of the Hamiltonian, $f$, utilized in this work are exclusive functions of the semi-major axis ratio $\alpha = a/a' < 1$, where $a'$ is the perturbing body's semi-major axis. The particular expansion of the gravitational potential we adopt here \citep{2000Icar..147..129E} utilizes Laplace coefficients, defined as 
\begin{align}
b_{\ell}^{(k)} = \frac{1}{\pi} \int_{0}^{2\pi} \frac{\cos(k \varphi)}{(1-2\alpha \cos(\varphi)+\alpha^2)^{\ell}}d \varphi
\end{align}
as well as their derivatives. Adopting the $\partial_{\alpha} = \partial/\partial \alpha$ notation for the differential operator, the expressions for the coefficients take on the following form.
\begin{align}
f_{e}^{(1)}  &= \frac{1}{8} \left(2 \alpha \partial_{\alpha} b_{1/2}^{(0)} +  \alpha^2 \partial_{\alpha}^2 b_{1/2}^{(0)} \right) \nonumber \\
f_{e}^{(2)}  &= \frac{1}{4} \left(2 b_{1/2}^{(1)} - 2 \alpha \partial_{\alpha} b_{1/2}^{(1)} - \alpha^2 \partial_{\alpha}^2 b_{1/2}^{(1)} \right) \nonumber \\
f_{e}^{(3)} &= \frac{1}{128} \left(4 \alpha^3 \partial_{\alpha}^3 b_{1/2}^{(0)} + \alpha^4 \partial_{\alpha}^4 b_{1/2}^{(0)} \right) \nonumber \\
f_{i}^{(1)}  &= -\frac{1}{2} \alpha b_{3/2}^{(1)}  \nonumber \\
f_{i}^{(2)}  &=  \alpha b_{3/2}^{(1)} \nonumber \\
f_{i}^{(3)} &= \frac{3}{16} \alpha^2 b_{5/2}^{(-2)} + \frac{3}{4} \alpha^2 b_{5/2}^{(0)} + \frac{3}{16} \alpha^2 b_{5/2}^{(2)} \nonumber \\
f_{ei} &= \frac{1}{16} \left(-2 \alpha ( b_{3/2}^{(-1)} + b_{3/2}^{(1)}) - 4 \alpha^2 (\partial_\alpha b_{3/2}^{(-1)} + \partial_\alpha b_{3/2}^{(1)}) - \alpha^3 (\partial_\alpha^2 b_{3/2}^{(-1)}+ \partial_\alpha^2 b_{3/2}^{(1)}) \right).
\end{align}

\section{Equations of Motion}
The equations of motion arising from Hamiltonians (\ref{HtwoDOF}) and (\ref{HthreeDOF}) as written in action-angle variables posses coordinate singularities at null actions and are thus unfavorable for practical use. Fortunately, this inconvenience can be remedied by a canonical change of variables. In terms of global cartesian coordinates \citep{Morbybook}
\begin{align}
\bar{x} &= \sqrt{2 \Phi} \cos(\phi) \ \ \ \ \ \ \ \ \ \ \bar{y} = \sqrt{2 \Phi} \sin(\phi) \nonumber, \\
\bar{w} &= \sqrt{2 \Psi} \cos(\psi) \ \ \ \ \ \ \ \ \ \bar{z} = \sqrt{2 \Psi} \sin(\psi), 
\label{cartcoords}
\end{align}
Hamiltonian (\ref{HthreeDOF}) takes on the following form:
\begin{align}
\tilde{\mathcal{H}} &= \left(F_{\rm{GR}} + \Feone + g_5 \right) \left(\frac{\bar{x}^2 + \bar{y}^2}{2} \right) + \Fethree \left(\frac{\bar{x}^2 + \bar{y}^2}{2} \right)^2 + \Fetwofive \left( \frac{\bar{x}}{\sqrt{2}} \right) + \left(\Fione + f_2 \right) \left(\frac{\bar{w}^2 + \bar{z}^2}{2} \right)+ \Fithree \left(\frac{\bar{w}^2 + \bar{z}^2}{2} \right)^2 \nonumber \\ &+ \Fitwo \left(\frac{w}{\sqrt{2}} \right) + \Fei \left(\frac{\bar{x}^2 + \bar{y}^2}{2} \right) \left(\frac{\bar{w}^2 + \bar{z}^2}{2} \right) + \Fetwotwo \left( \frac{\bar{x} \cos(\omega t + \Delta\beta_{2,5})}{\sqrt{2}} - \frac{\bar{y} \sin(\omega t + \Delta\beta_{2,5})}{\sqrt{2}} \right).
\label{HthreeDOFcart}
\end{align}
Note that in the above expression, we have adopted the notation used in section (4): $\omega = (g_2 - g_5)$ and $\Delta\beta_{2,5} = (\beta_2 - \beta_5)$.

Expression (\ref{HthreeDOFcart}) can be made more succinct by introducing complex canonical variables \citep{1966RvMP...38...36S} 
\begin{align}
\eta &=\frac{\bar{x}+\imath \bar{y}}{\sqrt{2}} \ \ \ \ \ \ \ \ \ \ \mu =\frac{\bar{w}+\imath \bar{z}}{\sqrt{2}},
\label{complexcoords}
\end{align}
where $\imath = \sqrt{-1}$. Accordingly, the Hamiltonian is rewritten as follows:
\begin{align}
\tilde{\mathcal{H}} &= \left(F_{\rm{GR}} + \Feone + g_5 \right) \left( \eta \eta^* \right) + \Fethree \left(\eta \eta^* \right)^2 + \Fetwofive \left( \frac{\eta + \eta^*}{2} \right) + \left(\Fione + f_2 \right) \left(\mu \mu^* \right)+ \Fithree \left(\mu \mu^* \right)^2 \nonumber \\ &+ \Fitwo \left( \frac{\mu + \mu^*}{2} \right) + \Fei \left( \eta \eta^* \right) \left( \mu \mu^* \right) + \Fetwotwo \left(\frac{\eta e^{\imath(\omega t + \Delta\beta_{2,5})} + \eta^* e^{-\imath(\omega t + \Delta\beta_{2,5})}}{2} \right).
\label{Hcomplex}
\end{align}
An additional advantage of introducing complex coordinates (\ref{complexcoords}) is that instead of integrating four real equations of motion, one needs only to integrate two complex ones. 

In complex form, Hamilton's equations become \citep{1966RvMP...38...36S}: 
\begin{align}
\frac{d\eta}{dt} &=\imath \frac{\partial \tilde{\mathcal{H}}}{\partial \eta^*} \ \ \ \ \ \ \ \ \ \ \frac{d\mu}{dt} =\imath \frac{\partial \tilde{\mathcal{H}}}{\partial \mu^*}.
\label{Hamiltonseqns}
\end{align}
Correspoindingly, the equations of motion read:
\begin{align}
\frac{d\eta}{dt} &= \left(F_{\rm{GR}} + \Feone + g_5 \right)\eta + 2 \Fethree \eta |\eta |^2 + \frac{1}{2} \Fetwofive + \Fei \eta |\mu |^2  + \Fetwotwo \left(e^{-\imath(\omega t + \Delta\beta_{2,5})} \right) \nonumber \\
\frac{d\mu}{dt} &= \left(\Fione + f_2 \right) \mu+ 2 \Fithree \mu | \mu |^2 + \frac{1}{2} \Fitwo + \Fei \mu |\eta |^2.
\label{EOM}
\end{align}
Naturally, the above expressions can be reduced to the equations of motion of the autonomous system (\ref{HtwoDOF}) by setting $\Fetwotwo = 0$. 

\end{appendix}


\begin{thebibliography} 

\bibitem[Adams(1846)]{Adams1846} Adams, J.~C.\ 1846, \mnras, 7, 149
\bibitem[Adams \& Laughlin(2006)]{2006IJMPD..15.2133A} Adams, F.~C., \& Laughlin, G.\ 2006, International Journal of Modern Physics D, 15, 2133 

\bibitem[Applegate et al.(1986)]{1986AJ.....92..176A} Applegate, J.~H., Douglas, M.~R., Gursel, Y., Sussman, G.~J., \& Wisdom, J.\ 1986, \aj, 92, 176 

\bibitem[Arnold(1961)]{Arnold1961} Arnold, V. I., 1961, Russ. Math. Surv. 18, 85
\bibitem[Arnold(1963)]{Arnold1963} Arnold, V. I., 1963, Uspehi Mat. Nauk 18, 13
\bibitem[Arnold(1964)]{Arnold1964} Arnold, V. I., 1964, Soviet Mathematics 5, 581
\bibitem[Arnold(1978)]{1978mmcm.book.....A} Arnold, V. I., 1978, Mathematical Methods of Classical Mechanics, (New York: Springer)

\bibitem[Batygin \& Laughlin(2008)]{BatLaugh} Batygin, K., \& Laughlin, G.\ 2008, \apj, 683, 1207 
\bibitem[Batygin(2010)]{2010NIMPA.618...37B} Batygin, Y.~K.\ 2010, Nuclear Instruments and Methods in Physics Research A, 618, 37 
\bibitem[Batygin \& Brown(2010)]{2010ApJ...716.1323B} Batygin, K., \& Brown, M.~E.\ 2010, \apj, 716, 1323 
\bibitem[Batygin et al.(2011)]{2011ApJ...738...13B} Batygin, K., Brown, M.~E., \& Fraser, W.~C.\ 2011, \apj, 738, 13 
\bibitem[Batygin \& Morbidelli(2011)]{2011CeMDA.111..219B} Batygin, K., \& Morbidelli, A.\ 2011, Celestial Mechanics and Dynamical Astronomy, 111, 219
\bibitem[Batygin \& Morbidelli(2013)]{2013A&A...556A..28B} Batygin, K., \& Morbidelli, A.\ 2013, \aap, 556, A28 


\bibitem[Bou{\'e} et al.(2012)]{Boue} Bou{\'e}, G., Laskar, J., \& Farago, F.\ 2012, \aap, 548, A43 

\bibitem[Bretagnon(1974)]{Bretagnon} Bretagnon, P.\ 1974, \aap, 30, 141 
\bibitem[Brouwer \& Van Woerkom(1950)]{WorkemBurkem} Brouwer, D., \& Van Woerkom, A. J. J.\ 1950, Astr. Papers Amer. Ephemeris 13, Part II
\bibitem[Brouwer \& Clemence(1961)]{1961mcm..book.....B} Brouwer, D., \& Clemence, G.~M.\ 1961, Methods of Celestial Mechanics (New York: Academic Press)

\bibitem[Carpino et al.(1987)]{1987A&A...181..182C} Carpino, M., Milani, A., \& Nobili, A.~M.\ 1987, \aap, 181, 182 

\bibitem[Cary et al.(1986)]{1986PhRvA..34.4256C} Cary, J.~R., Escande, D.~F., \& Tennyson, J.~L.\ 1986, \pra, 34, 4256 

\bibitem[Celletti \& Chierchia(1997)]{1997CMaPh.186..413C} Celletti, A., \& Chierchia, L.\ 1997, Communications in Mathematical Physics, 186, 413 

\bibitem[Chatterjee et al.(2008)]{Chaterjee08} Chatterjee, S., Ford, E.~B., Matsumura, S., \& Rasio, F.~A.\ 2008, \apj, 686, 580 

\bibitem[Chirikov(1959)]{Chirikov1959} Chirikov, B.~V.\ 1959, Soviet Physics Doklady, 4, 390 
\bibitem[Chirikov et al.(1971)]{1971JSP.....3..307C} Chirikov, B.~V., Keil, E., \& Sessler, A.~M.\ 1971, Journal of Statistical Physics, 3, 307
\bibitem[Chirikov(1979)]{Chirikov1979} Chirikov, B.~V.\ 1979, \physrep, 52, 263 

\bibitem[Cincotta \& Sim{\'o}(2000)]{2000A&AS..147..205C} Cincotta, P.~M., \& Sim{\'o}, C.\ 2000, \aaps, 147, 205 

\bibitem[Cumming(2011)]{Cumming} Cumming, A.\ 2011, Exoplanets, edited by S.~Seager.~ (Tucson, AZ: University of Arizona Press)
 
\bibitem[Deck et al.(2013)]{2013ApJ...774..129D} Deck, K.~M., Payne, M., \& Holman, M.~J.\ 2013, \apj, 774, 129

\bibitem[Deprit(1969)]{1969CeMec...1...12D} Deprit, A.\ 1969, Celestial Mechanics, 1, 12 

\bibitem[Ellis \& Murray(2000)]{2000Icar..147..129E} Ellis, K.~M., \& Murray, C.~D.\ 2000, \icarus, 147, 129

\bibitem[Escande(1985)]{1985PhR...121..165E} Escande, D.~F.\ 1985, \physrep, 121, 165 

\bibitem[Ford \& Rasio(2008)]{FordRasio08} Ford, E.~B., \& Rasio, F.~A.\ 2008, \apj, 686, 621

\bibitem[Gauss(1809)]{Gauss1809} Gauss, C. F., 1809, Werke Vol. 3, 331, Gottinger, Dieterich (1866).

\bibitem[Gluckstern(1994)]{1994PhRvL..73.1247G} Gluckstern, R.~L.\ 1994, Physical Review Letters, 73, 1247 

\bibitem[Goldstein(1950)]{Goldstein} Goldstein, H.\ 1950, Classical Mechanics, (Mass.: Addison-Wesley)  

\bibitem[Grimmett \& Stirzaker(2001)]{probabilitybook} Grimmett, G. \& Stirzaker, D.\ 2001, Probability and Random Processes (Oxford, UK: Oxford University Press)

\bibitem[Henrard(1982)]{1982amdc.proc..153H} Henrard, J.\ 1982, NATO ASIC Proc.~82: Applications of Modern Dynamics to Celestial Mechanics and 
Astrodynamics, 153 
\bibitem[Henrard \& Lamaitre(1983)]{1983CeMec..30..197H} Henrard, J., \& Lamaitre, A.\ 1983, Celestial Mechanics, 30, 197 
\bibitem[Henrard \& Caranicolas(1990)]{1990CeMDA..47...99H} Henrard, J., \& Caranicolas, N.~D.\ 1990, Celestial Mechanics and Dynamical Astronomy, 47, 99
\bibitem[Henrard \& Henrard(1991)]{1991PhyD...54..135H} Henrard, J., \& Henrard, M.\ 1991, Physica D Nonlinear Phenomena, 54, 135 

\bibitem[H\'enon(1966)]{Henon1966} H\'enon, M.\ 1966, Bulletin Astronomique, 3, 49 

\bibitem[Holman \& Wisdom(1993)]{1993AJ....105.1987H} Holman, M.~J., \& Wisdom, J.\ 1993, \aj, 105, 1987 
\bibitem[Holman \& Murray(1996)]{HolmanMurray1996} Holman, M.~J., \& Murray, N.~W.\ 1996, \aj, 112, 1278 

\bibitem[Hori(1966)]{1966PASJ...18..287H} Hori, G.\ 1966, \pasj, 18, 287 

\bibitem[Ito \& Tanikawa(2002)]{2002MNRAS.336..483I} Ito, T., \& Tanikawa, K.\ 2002, \mnras, 336, 483

\bibitem[Juri{\'c} \& Tremaine(2008)]{JuricTremaine} Juri{\'c}, M., \& Tremaine, S.\ 2008, \apj, 686, 603 

\bibitem[Klebaner(2012)]{Stochastic} Klebaner, F. C., 1998, Introduction to Stochastic Calculus with Applications (World Scientific Publishing)

\bibitem[Kolmogorov(1954)]{Kolmogorov} Kolmogorov, A. N., 1954, Dokl. Akad. Nauk SSSR 98, 527.

\bibitem[Lagrange(1778)]{Lagrange} Lagrange, J. L., 1778, M\'emoires de l'Acad\'emie des Sciences de Paris, ann\'ee 1774.

\bibitem[Landau(1937)]{1937PhRv...52.1251L} Landau, L.\ 1937, Physical Review, 52, 1251 

\bibitem[Laplace(1772)]{Laplace72} Laplace, P. S., 1772, Oeuvres compl\'etes, 9, 325, Paris, Gauthier-Villars (1895)
\bibitem[Laplace(1775)]{Laplace75} Laplace, P. S., 1775, M\'emoires de l'Acad\'emie des Sciences de Paris, ann\'ee 1772.

\bibitem[Laskar(1989)]{Laskar89} Laskar, J.\ 1989, \nat, 338, 237 
\bibitem[Laskar(1990)]{Laskar90} Laskar, J.\ 1990, \icarus, 88, 266 
\bibitem[Laskar(1994)]{Laskar94} Laskar, J.\ 1994, \aap, 287, L9 
\bibitem[Laskar \& Robutel(1995)]{1995CeMDA..62..193L} Laskar, J., \& Robutel, P.\ 1995, Celestial Mechanics and Dynamical Astronomy, 62, 193 
\bibitem[Laskar(1996)]{Laskar96} Laskar, J.\ 1996, Celestial Mechanics and Dynamical Astronomy, 64, 115 
\bibitem[Laskar(2008)]{Laskar08} Laskar, J.\ 2008, \icarus, 196, 1 
\bibitem[Laskar \& Gastineau(2009)]{Laskar09} Laskar, J., \& Gastineau, M.\ 2009, \nat, 459, 817 
\bibitem[Laskar(2012)]{LaskarPoincare} Laskar, J.\ 2012, arXiv:1209.5996 

\bibitem[Lecar et al.(1992)]{1992AJ....104.1230L} Lecar, M., Franklin, F., \& Murison, M.\ 1992, \aj, 104, 1230

\bibitem[Lega et al.(2013)]{2013MNRAS.431.3494L} Lega, E., Morbidelli, A., \& Nesvorn{\'y}, D.\ 2013, \mnras, 431, 3494 

\bibitem[Leverrier(1855)]{LeVerrier1855} Leverrier, U. J.\ 1855, Annales de llObservatoire de Paris, 1, 258

\bibitem[Levison et al.(2008)]{2008Icar..196..258L} Levison, H.~F., Morbidelli, A., Van Laerhoven, C., Gomes, R., \& Tsiganis, K.\ 2008, \icarus, 196, 258 
\bibitem[Levison et al.(2011)]{2011AJ....142..152L} Levison, H.~F., Morbidelli, A., Tsiganis, K., Nesvorn{\'y}, D., \& Gomes, R.\ 2011, \aj, 142, 152

\bibitem[Lichtenberg \& Lieberman(1983)]{Yellowbook} Lichtenberg, A.~J., \& Lieberman, M.~A.\ 1983, Regular and Chaotic Dynamics (Applied Mathematical Sciences), (New York: Springer)

\bibitem[Lithwick \& Wu(2011)]{LithWu11} Lithwick, Y., \& Wu, Y.\ 2011, \apj, 739, 31 

\bibitem[Locatelli \& Giorgilli(2000)]{2000CeMDA..78...47L} Locatelli, U., \& Giorgilli, A.\ 2000, Celestial Mechanics and Dynamical Astronomy, 78, 47 


\bibitem[Michtchenko \& Malhotra(2004)]{2004Icar..168..237M} Michtchenko, T.~A., \& Malhotra, R.\ 2004, \icarus, 168, 237 

\bibitem[Migaszewski \& Go{\'z}dziewski(2009a)]{2009MNRAS.392....2M} Migaszewski, C., \& Go{\'z}dziewski, K.\ 2009a, \mnras, 392, 2 
\bibitem[Migaszewski \& Go{\'z}dziewski(2009b)]{2009MNRAS.395.1777M} Migaszewski, C., \& Go{\'z}dziewski, K.\ 2009b, \mnras, 395, 1777

\bibitem[Morbidelli \& Giorgilli(1990a)]{Morby90a} Morbidelli, A., \& Giorgilli, A.\ 1990a, Celestial Mechanics and Dynamical Astronomy, 47, 145
\bibitem[Morbidelli \& Giorgilli(1990b)]{Morby90b} Morbidelli, A., \& Giorgilli, A.\ 1990b, Celestial Mechanics and Dynamical Astronomy, 47, 173  
\bibitem[Morbidelli \& Henrard(1991a)]{Morby91a} Morbidelli, A., \& Henrard, J.\ 1991, Celestial Mechanics and Dynamical Astronomy, 51, 131
\bibitem[Morbidelli \& Henrard(1991b)]{Morby91b} Morbidelli, A., \& Henrard, J.\ 1991, Celestial Mechanics and Dynamical Astronomy, 51, 169 

\bibitem[Morbidelli(1993)]{1993CeMDA..55..101M} Morbidelli, A.\ 1993, Celestial Mechanics and Dynamical Astronomy, 55, 101 

\bibitem[Morbidelli \& Giorgilli(1993)]{1993CeMDA..55..131M} Morbidelli, A., \& Giorgilli, A.\ 1993, Celestial Mechanics and Dynamical Astronomy, 55, 131 

\bibitem[Morbidelli \& Giorgilli(1995)]{1995JSP....78.1607M} Morbidelli, A., \& Giorgilli, A.\ 1995, Journal of Statistical Physics, 78, 1607 

\bibitem[Morbidelli(2002)]{Morbybook} Morbidelli, A.\ 2002, Modern Celestial Mechanics: Aspects of Solar System Dynamics (Taylor \& Francis)

\bibitem[Moser(1962)]{Moser} Moser, J., 1962,  Nachr. Akad. Wiss. Gšttingen Math.-Phys. Kl. 2, 1

\bibitem[Murray et al.(1985)]{1985PhRvA..32.2413M} Murray, N.~W., Lieberman, M.~A., \& Lichtenberg, A.~J.\ 1985, \pra, 32, 2413 
\bibitem[Murray \& Holman(1997)]{MurrayHolman97} Murray, N., \& Holman, M.\ 1997, \aj, 114, 1246 
\bibitem[Murray \& Holman(1999)]{MurrayHolman99} Murray, N., \& Holman, M.\ 1999, Science, 283, 1877 
\bibitem[Murray \& Holman(2001)]{2001Natur.410..773M} Murray, N., \& Holman, M.\ 2001, \nat, 410, 773

\bibitem[Murray \& Dermott(1999)]{MD99} Murray, C.~D., \& Dermott, S.~F.\ 1999, Solar System Dynamics (Cambridge: Cambridge Univ. Press) 

\bibitem[Neishtadt(1984)]{1984PriMM..48..197N} Neishtadt, A.~I.\ 1984, Prikladnaia Matematika i Mekhanika, 48, 197

\bibitem[Nekhoroshev(1977)]{1977RuMaS..32....1N} Nekhoroshev, N.~N.\ 1977, Russian Mathematical Surveys, 32, 1 

\bibitem[Nesvorn{\'y} \& Morbidelli(2012)]{2012AJ....144..117N} Nesvorn{\'y}, D., \& Morbidelli, A.\ 2012, \aj, 144, 117 
\bibitem[Nesvorn{\'y} \& Morbidelli(1998a)]{DavidMorbya} Nesvorn{\'y}, D., \& Morbidelli, A.\ 1998, \aj, 116, 3029 
\bibitem[Nesvorn{\'y} \& Morbidelli(1998b)]{DavidMorbyb} Nesvorn{\'y}, D., \& Morbidelli, A.\ 1998, Celestial Mechanics and Dynamical Astronomy, 71, 243 

\bibitem[Nesvorn{\'y} \& Roig(2000)]{2000Icar..148..282N} Nesvorn{\'y}, D., \& Roig, F.\ 2000, \icarus, 148, 282 
\bibitem[Nesvorn{\'y} \& Roig(2001)]{2001Icar..150..104N} Nesvorn{\'y}, D., \& Roig, F.\ 2001, \icarus, 150, 104 

\bibitem[Poincare(1892)]{Poincare92} Poincare, H.\ 1892,\ Les M\'ethodes Nouvelles de la M\'ecanique C\'eleste, Vol. 1-3.\ Paris, Gauthier-Villars et fils (1892-99)

\bibitem[Poisson(1809)]{Poisson} Poisson, 1809, Journal de l'Ecole Polytechnique, Cahier XV, 8, 1

\bibitem[Quinn et al.(1991)]{Quinn1991} Quinn, T.~R., Tremaine, S., \& Duncan, M.\ 1991, \aj, 101, 2287

\bibitem[Rasio \& Ford(1996)]{RasioFord96} Rasio, F.~A., \& Ford, E.~B.\ 1996, Science, 274, 954 
\bibitem[Raymond et al.(2009a)]{Raymond09a} Raymond, S.~N., Barnes, R., Veras, D., et al.\ 2009a, \apjl, 696, L98 
\bibitem[Raymond et al.(2009b)]{Raymond09b} Raymond, S.~N., Armitage, P.~J., \& Gorelick, N.\ 2009b, \apjl, 699, L88 

\bibitem[Sidlichovsky(1990)]{Sid1990} Sidlichovsky, M.\ 1990, Celestial Mechanics and Dynamical Astronomy, 49, 177

\bibitem[Strocchi(1966)]{1966RvMP...38...36S} Strocchi, F.\ 1966, Reviews of Modern Physics, 38, 36 

\bibitem[Sussman \& Wisdom(1988)]{SussmanWisdom1988} Sussman, G.~J., \& Wisdom, J.\ 1988, Science, 241, 433 
\bibitem[Sussman \& Wisdom(1992)]{SussmanWisdom1992} Sussman, G.~J., \& Wisdom, J.\ 1992, Science, 257, 56 

\bibitem[Tsiganis et al.(2005)]{2005Natur.435..459T} Tsiganis, K., Gomes, R., Morbidelli, A., \& Levison, H.~F.\ 2005, \nat, 435, 459

\bibitem[Veras \& Armitage(2007)]{2007ApJ...661.1311V} Veras, D., \& Armitage, P.~J.\ 2007, \apj, 661, 1311 

\bibitem[Wang \& Uhlenbeck(1945)]{1945RvMP...17..323W} Wang, M.~C., \& Uhlenbeck, G.~E.\ 1945, Reviews of Modern Physics, 17, 323

\bibitem[Wisdom(1980)]{Wisdom80} Wisdom, J.\ 1980, \aj, 85, 1122 
\bibitem[Wisdom(1983)]{Wisdom83} Wisdom, J.\ 1983, \icarus, 56, 51 
\bibitem[Wisdom(1986)]{1986CeMec..38..175W} Wisdom, J.\ 1986, Celestial Mechanics, 38, 175 

\bibitem[Wu \& Lithwick(2011)]{2011ApJ...735..109W} Wu, Y., \& Lithwick, Y.\ 2011, \apj, 735, 109 

%Laplace, P.-S., 1775, M?emoire sur les solutions particuli`eres des ?equations diff?erentielles et sur les in?egalit?es s?eculaires des plan`etes. M?emoires de lÕAcad?emie des Sciences de Paris, ann?ee 1772, publi?e en 1775, Îuvres, t. VIII, p. 325
%Poincare, H. : 1892-1899, Les M?ethodes Nouvelles de la M?ecanique C?eleste, tomes I-III, Gauthier Villard, Paris, reprinted by Blanchard, 1987
%\bibitem[Poincare(1892)]{1892QB351.P75......} Poincare, H.\ 1892, Paris, Gauthier-Villars et fils, 1892-99.,
%Poincar?e, H. : 1897, Sur la stabilit?e du Syst`eme Solaire, Annuaire du Bureau des Longitudes pour lÕan 1898, Paris, Gauthier-Villars, B1-B16
%\bibitem[Leverrier(1860)]{1860MNRAS..20..303L} Leverrier, U.~J.\ 1860, \mnras, 20, 303 

\end{thebibliography}
\end{document}